\documentclass[reprint,prl,aps,superscriptaddress,twocolumn, 10pt]{revtex4-1}
\usepackage{amsmath}
\usepackage[english]{babel}
\usepackage{graphicx}
\usepackage{color}
\usepackage[dvipsnames]{xcolor}
\usepackage{sidecap}
\usepackage{textcomp}

\usepackage[colorlinks,
            citecolor = blue, 
						linkcolor = blue,
						urlcolor  = blue,
						anchorcolor = blue,
						breaklinks = true
						]{hyperref}

\usepackage{soul}
\usepackage{ amssymb }

\newif\ifdraft
\drafttrue

\newcommand{\eV}{\ensuremath{\mathrm{\,eV}}}

\newcommand{\tesla}{\ensuremath{\mathrm{\,T}}}
\newcommand{\nW}{\ensuremath{\mathrm{\,nW}}}
\newcommand{\nm}{\ensuremath{\mathrm{\,nm}}}

\newcommand{\kelvin}{\ensuremath{\mathrm{\,K}}}
\newcommand{\mkelvin}{\ensuremath{\mathrm{\,mK}}}
\newcommand{\volt}{\ensuremath{\mathrm{\,V}}}
\newcommand{\mvolt}{\ensuremath{\mathrm{\,mV}}}

\newcommand{\vtg}{\ensuremath{V_\mathrm{TG}}}
\newcommand{\vbg}{\ensuremath{V_\mathrm{BG}}}

\newcommand{\MoSe}{\ensuremath{\mathrm{MoSe}_2}}

\newcommand{\textapprox}{\raisebox{0.5ex}{\texttildelow}}
\newcommand{\figref}[1]{Fig.\,\ref{#1}}

\def \ETH{Institute for Quantum Electronics, ETH Z\"urich, CH-8093 Z\"urich, Switzerland}
\def \NIMSRCFM{Research Center for Functional Materials, National Institute for Materials Science, Tsukuba, Ibaraki 305-0044, Japan}
\def \NIMSICMN{International Center for Materials Nanoarchitectonics, National Institute for Materials Science, Tsukuba, Ibaraki 305-0044, Japan}

\def \RIKEN{Center for Emergent Matter Science, RIKEN, Wako, Saitama 351-0198, Japan}

\begin{document}


\title{Optical sensing of fractional quantum Hall effect in graphene} 

\author{A. Popert}
\affiliation{\ETH}

\author{Y. Shimazaki}
\affiliation{\ETH}
\affiliation{\RIKEN}

\author{M. Kroner}
\affiliation{\ETH}

\author{K.~Watanabe}
\affiliation{\NIMSRCFM}

\author{T.~Taniguchi}
\affiliation{\NIMSICMN}

\author{A. Imamo\u{g}lu}
\affiliation{\ETH}

\author{T. Smole\'nski}
\affiliation{\ETH}

\begin{abstract}
Graphene and its van der Waals (vdW) heterostructures provide a unique and versatile playground for explorations of strongly correlated electronic phases, ranging from unconventional fractional quantum Hall (FQH) states in a monolayer system to a plethora of superconducting and insulating states in twisted bilayers. However, the access to those fascinating phases has been thus far entirely restricted to transport techniques, due to the lack of a robust energy bandgap that makes graphene hard to access optically. Here we demonstrate an all-optical, non-invasive spectroscopic tool for probing electronic correlations in graphene using excited Rydberg excitons in an adjacent transition metal dichalcogenide monolayer. Due to their large Bohr radii, Rydberg states are highly susceptible to the compressibility of graphene electrons, allowing us to detect the formation of odd-denominator FQH states at high magnetic fields. Owing to its sub-micron spatial resolution, the technique we demonstrate circumvents spatial inhomogeneities in vdW structures, and paves the way for optical studies of correlated states in twisted bilayer graphene and other optically inactive atomically-thin materials.
\end{abstract}

\maketitle

Since their first mechanical exfoliation from a bulk graphite crystal~\cite{Novoselov_Science_2004,Novoselov_Nature_2005,Geim_Novoselov_NatMater}, graphene monolayers have emerged as a fascinating and promising experimental platform for explorations of two-dimensional electron systems (2DES). Owing to linear band dispersion, graphene subjected to an external magnetic field $B$ features an unusual sequence of anharmonically spaced Landau levels (LLs)~\cite{DasSarma2011}. Strong quenching of kinetic energy together with reduced dielectric screening compared to conventional GaAs-based 2DES markedly enhances the stability of fractional quantum Hall (FQH) states --- correlated electronic states arising in partially filled LLs due to electron-electron interactions~\cite{Du2009,Bolotin2009,Zeng2019}. FQH states typically emerge at odd-denominator fractional filling factors \mbox{$\nu=n/(2pn\pm1)$} for integer $n$, $p$~\cite{Tsui_PRL_1982,Laughlin_PRL_1983,FQH_review_1999}. Advances in fabrication of graphene-based devices have further enabled the observation of even more exotic correlated phases, including even-denominator FQH states~\cite{Li2017,Zibrov2018}, fractal FQH states originating from a superlattice moir{\'e} potential in angle-aligned graphene-hexagonal boron-nitride (hBN) heterostructures~\cite{Nature_Ponomarenko_2013,Dean2013,Yu_NatPhys_2011}, as well as superconducting and correlated Chern insulator phases in magic-angle twisted bilayer graphene (MATBG)~\cite{Cao2018,Cao2018a,Uri2020,Stepanov2020,Liu2021}.

Most of the prior investigations of strong electronic correlations in graphene have been limited to transport experiments. This is due to the lack of a robust bandgap, which hinders usage of interband optical transitions to probe the formation of correlated states in a similar way as has been done for GaAs-based quantum wells~\cite{Heiman_PRL_1988,Turberfield_PRL_1990,Yusa2001,Byszewski2006}. This motivates the question: can such states in graphene be investigated by purely optical means? Optical spectroscopy with its high spatial resolution allows to circumvent both the influence of device edges and disorder-related effects arising from inevitable spatial inhomogeneities introduced during assembly of van der Waals heterostructures. Furthermore, optical probes offer ultra-fast access to correlated electronic states. They may therefore enable explorations of elusive correlated phases being thus far inaccessible in transport techniques.

\begin{figure*}
    \includegraphics{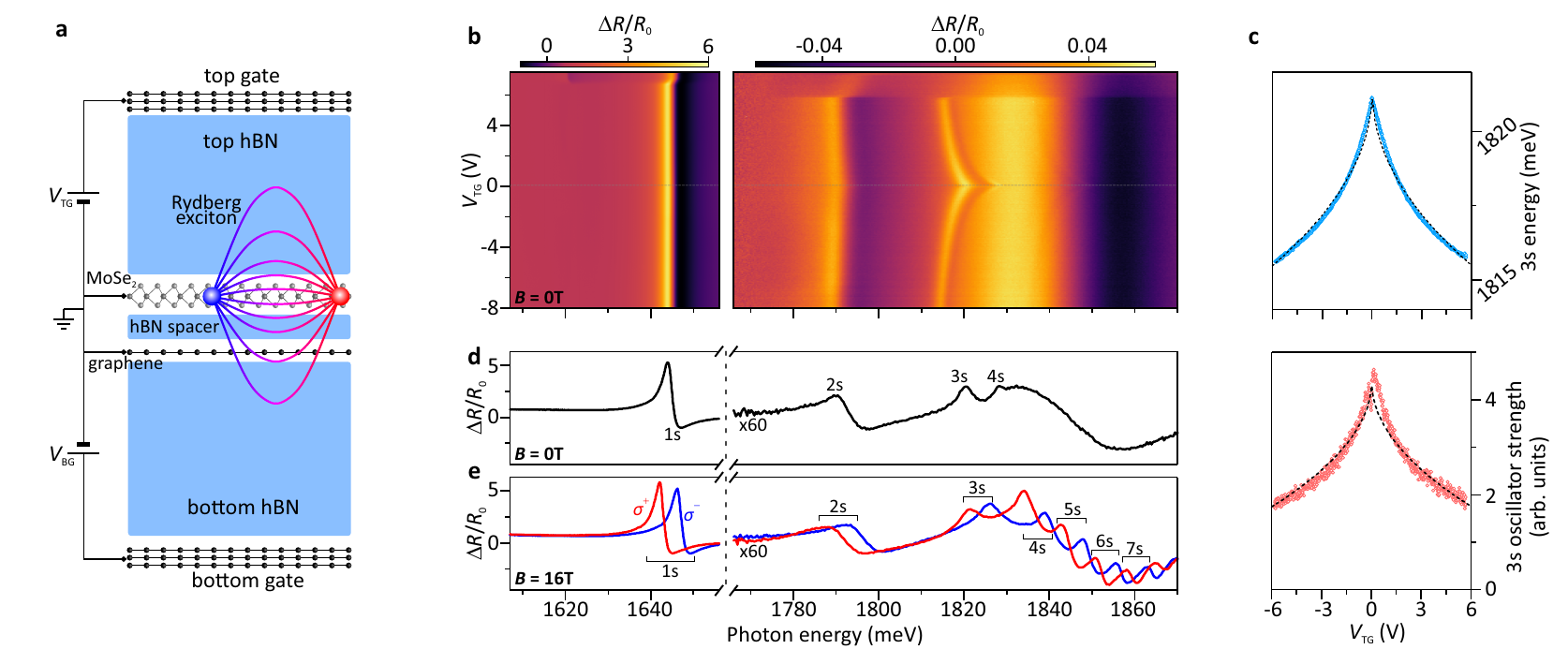}
	\caption{{\bf Sensing the compressibility of graphene electrons with Rydberg excitons.} ({\bf a})~Cartoon illustrating the layer structure of of the sample region investigated in the main text. The MoSe$_2$ monolayer is separated from a proximal graphene layer by a 5ML-thick hBN flake.~({\bf b})~\vtg-evolution of the zero-field $\Delta R/R_0$ spectra acquired at $\vbg\approx0\volt$ and at $T\approx4$~K in the spectral range of the ground exciton (left) and the excited Rydberg states (right). The graphene charge-neutrality point at $\vtg\approx0\volt$ is marked by a horizontal dotted line. The onset of \MoSe\ monolayer doping occurs at $\vtg\approx6\volt$.~({\bf c}) Voltage-dependent energy (top) and oscillator strength (bottom) of the 3s exciton obtained by fitting its lineshape with a dispersive Lorentzian spectral profile. The dashed lines mark the fitted dependencies of the form $\propto1-a\sqrt{|n_e|}$ (where $a$ is a fitting parameter) that correspond to the reduction of both quantities being proportional to the quantum capacitance $C_Q\propto\sqrt{|n_e|}$ that in turn determines graphene compressibility $\kappa\propto C_Q/n_e^2$. {\bf d} Line cut through the maps in {\bf b} at $\vtg\approx0$\volt.~({\bf e}) Reflectance contrast spectra measured on the same spot at $B=16\tesla$ in two circular polarizations.\label{fig:Fig1}} 
\end{figure*}

Here, we demonstrate all-optical detection of the FQH effect in graphene with the use of a proximal transition metal dichalcogenide (TMD) monolayer. The utilized technique exploits the fact that Coulomb interactions between charge carriers in a TMD monolayer are highly susceptible to the dielectric environment~\cite{Raja2019,Raja2017}. This allows to use excited Rydberg states of a TMD exciton as quantum proximity sensors for the compressibility of a 2DES in an adjacent layer, as first shown in Refs~\onlinecite{Raja2017,Xu2020,Xu2021}. Owing to graphene's almost perfect electron-hole symmetry~\cite{DasSarma2011} and low light absorption coefficient~\cite{Mak_PRL_2008}, this method does not affect the LL filling factor of the 2DES and hence constitutes a non-destructive probe for the correlations between the graphene electrons. The Rydberg exciton spectroscopy we implement is similarly sensitive to state-of-the-art transport techniques, but is more versatile as it permits to study FQH effect in simple device geometries  with high spatial resolution and without requiring electrical currents. 

The device we investigate consists of a graphene flake separated from a MoSe$_2$ monolayer by a thin hBN spacer (see \figref{fig:Fig1}\textbf{a}). The spacer exhibits a spatially-dependent thickness, in particular featuring hBN terraces of 2 or 5 monolayers (MLs). Unless otherwise stated, we exclusively explore the region with a 5ML-thick spacer in the main text [see Supplementary Information (SI) for additional data sets obtained in other sample regions]. The graphene-hBN-MoSe$_2$ structure is encapsulated between 40-50\nm-thick hBN slabs and then embedded in top and bottom graphite layers, which enable us to control the carrier density in the 2DES by applying gate voltages (\vtg\ and \vbg). The sample is loaded into a dilution refrigerator equipped with fiber-based optical access and a heater allowing to reach base temperatures ranging from 80\mkelvin\ to 20\kelvin\ (see SI for further details). The optical response of the TMD monolayer is analyzed by means of reflectance contrast $\Delta R/R_0 = (R - R_0)/R_0$ spectroscopy, where $R$ denotes the white-light reflectance spectrum of the device, while $R_0$ corresponds to the background spectrum (see SI).

Figure~\ref{fig:Fig1}\textbf{b} shows the top-gate voltage (\vtg) evolution of the reflectance contrast spectrum of our device acquired at $T=4\kelvin$ and at zero magnetic field. Due to the sizeable energy offset between the graphene Dirac point and the \MoSe\ conduction band edge~\cite{Wilson2017}, electron doping of the TMD only occurs at $\vtg\gtrsim5\volt$. This is where we observe attractive and repulsive Fermi-polarons for the 1s exciton~\cite{Sidler2017,Efimkin2017}. In this work, we focus on the opposite regime $\vtg\lesssim5\volt$, where the TMD remains charge neutral as carriers are introduced exclusively to the graphene layer. 

Under these conditions, the energy of the \MoSe\ 1s exciton is almost completely independent of the changes in graphene doping density $n_e$. This is due to extremely tight binding of this exciton state~\cite{Chernikov2014a} that renders its optical transition insensitive to screening by graphene electrons. More specifically, the screening-induced decrease of the 1s binding energy ($\Delta E_{1\mathrm{s}}$) is perfectly compensated by a corresponding reduction of the quasiparticle bandgap ($\Delta E_g$) of the TMD monolayer~\cite{Semkat2009,Riis-Jensen2020,Gjerding2020,Xu2021}. Such a compensation does not hold for the excited Rydberg exciton states $n$s (with $n=2,3,$...), which feature larger Bohr radii and smaller binding energies~\cite{Goryca2019,Molas2019}. As a consequence, we observe a prominent red-shift of the Rydberg exciton energies $\Delta E_g-\Delta E_{n\mathrm{s}}<0$ which is particularly well-resolved for the 3s exciton. This energy shift is directly proportional to the quantum capacitance  $C_Q = e^2 \partial n_e / \partial E_F$ of graphene electrons which increases non-linearly with the doping density, scaling as $\sqrt{|n_e|}$ due to the linear dispersion of graphene~\cite{Novoselov_Nature_2005,Geim_Novoselov_NatMater,DasSarma2011}. In parallel to the reduction of $\Delta E_{n\mathrm{s}}$, the oscillator strength of the Rydberg state also decreases $\propto\sqrt{|n_e|}$, as can be seen in \figref{fig:Fig1}\textbf{c}. These findings demonstrate that both the energy and the oscillator strength of the Rydberg excitons can be used as quantitative probes of the electronic compressibility $\kappa\propto C_Q/n_e^2$ of proximal graphene.

\begin{figure*}[t]
	\includegraphics{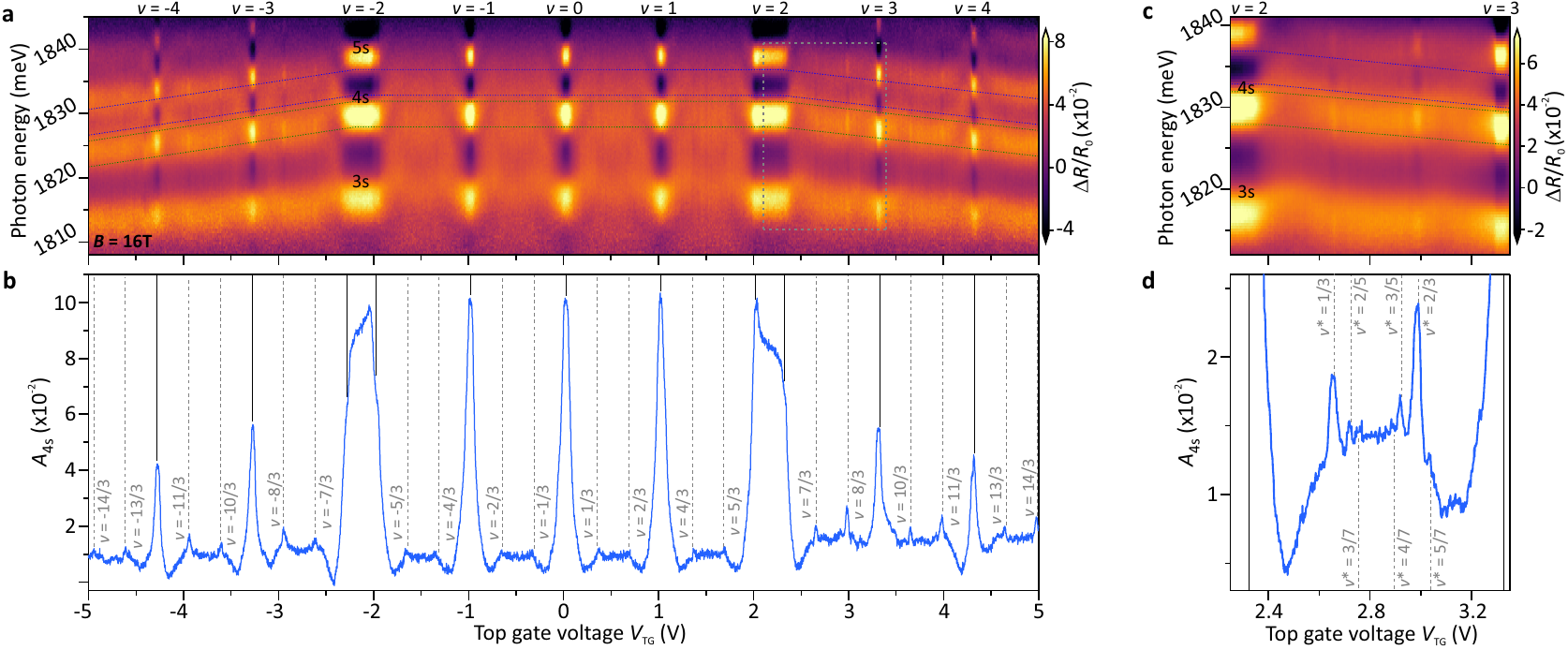}
	\caption{{\bf Optical probing of FQH states in graphene.} ({\bf a})~Top-gate-voltage evolution of the reflectance contrast spectra measured in the spectral range of the excited Rydberg exciton states at $B=16 \tesla$, $T=80 \mkelvin$ and for $\vbg=0 \volt$.~({\bf b})~\vtg-dependent amplitude of the 4s exciton determined as a difference between $\Delta R/R_0$ averaged over two 4-meV-wide spectral windows around the 4s resonance (marked by blue and green dashed lines in {\bf a}). The vertical solid (dashed) lines mark the subsequent integer (fractional) filling factors of graphene LLs.~({\bf c})~$\Delta R/R_0$ spectra acquired in a narrower voltage range corresponding to $2\lesssim\nu\lesssim3$ (indicated by the dotted rectangle in {\bf a}) with improved signal-to-noise ratio due to averaging over multiple measurements.~({\bf d})~The \vtg-dependent amplitude of the 4s exciton determined in a similar way as in {\bf b} based on the spectra in {\bf c}. The amplitude exhibits several prominent maxima revealing a multitude of FQH states with denominators of $3$, $5$ and $7$ (as indicated by the vertical dashed lines).\label{fig:Fig2}}
\end{figure*}

In order to exploit Rydberg exciton spectroscopy to study electronic correlations, we apply a strong external magnetic field $B=16\tesla$ that leads to the formation of Landau levels and thereby quenches the kinetic energy of graphene electrons. Concurrent quantization of motion of the TMD electrons and holes provides additional confinement to the Rydberg excitons. As a consequence, Rydberg excitons at $B=16\tesla$ are considerably stronger compared to their zero-field counterparts (see Figs~\ref{fig:Fig1}{\bf d,e}), which in turn improves the overall sensitivity of our sensing scheme. These excitons also remain bound for markedly higher principle quantum numbers exceeding $n\approx5$, for which the energy splitting between consecutive Rydberg states approaches the sum of electron and hole cyclotron energies in the TMD monolayer~\cite{Goryca2019,Molas2019}.

\begin{figure*}[t]
	\includegraphics{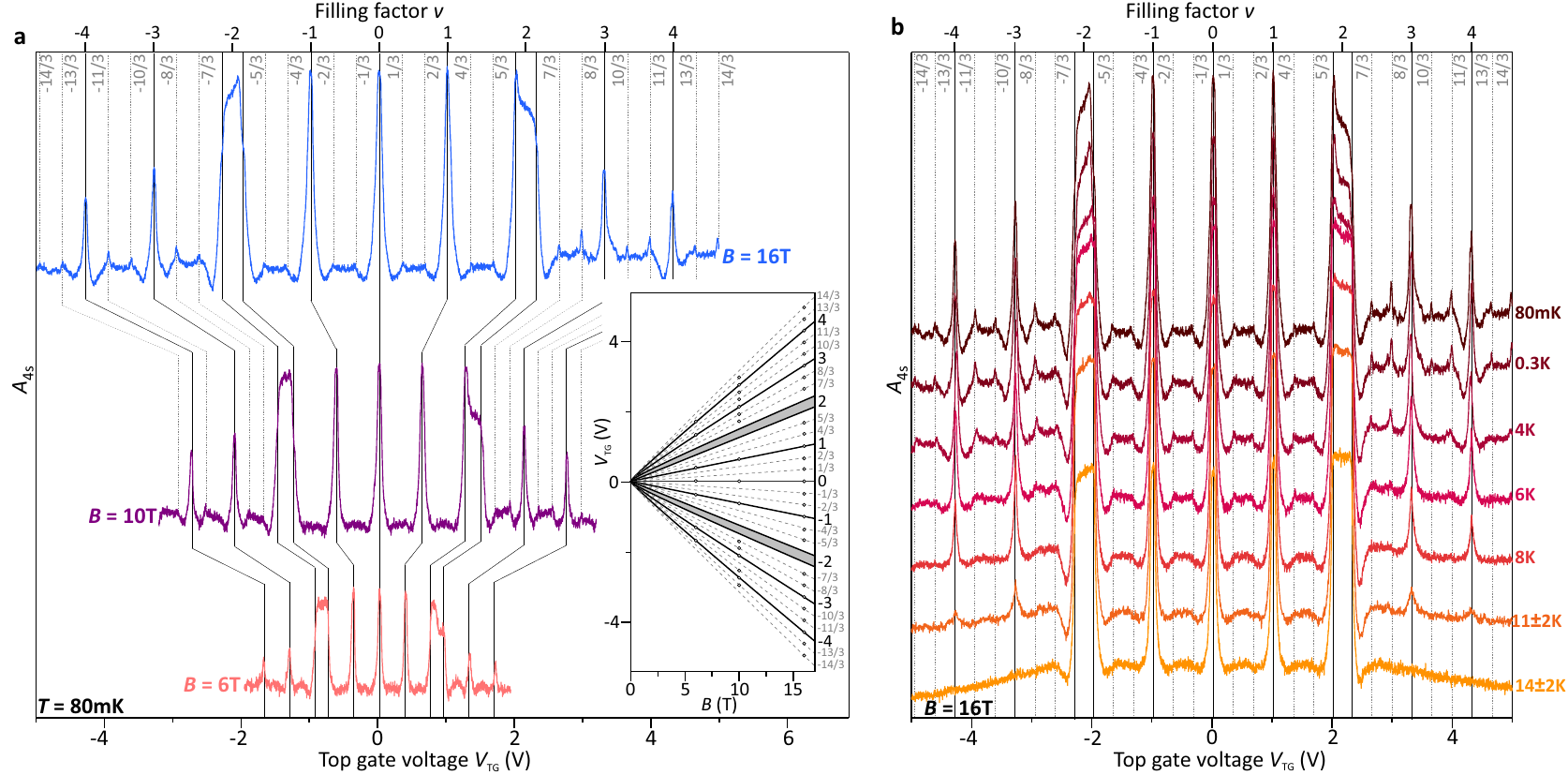}
	\caption{{\bf Magnetic field and temperature dependence of FQH states in graphene.} ({\bf a})~\vtg-dependent amplitude of the 4s exciton acquired at $T=80 \mkelvin$, $\vbg=0 \volt$, but at different magnetic fields of $B=16 \tesla$, $10 \tesla$ and $6 \tesla$ (the curves are vertically offset for clarity). The vertical solid (dashed) lines indicate integer (fractional) filling factors. Inset: top gate voltages corresponding to subsequent integer and fractional filling factors determined as a function of the magnetic field by fitting the positions of the 4s exciton amplitude maxima with phenomenological, Gaussian profiles (see SI for details). The set of solid and dashed lines represents the fit of the data with dependencies described in the text that form a LL fan chart.~({\bf b})~\vtg-dependent amplitude of the 4s exciton acquired at $B=16\tesla$, $\vbg=0 \volt$, but for different temperatures of sample (as indicated; the curves are vertically offset for clarity).\label{fig:Fig3}}
\end{figure*}

Under the influence of high $B$-field the Rydberg states exhibit a qualitatively different response to changes in graphene doping density. We show this in \figref{fig:Fig2}{\bf a} which displays \vtg-dependent reflectance contrast spectra in the 3s--6s energy range. The spectra were acquired at $T=80$~mK using low white-light power \textapprox 9\nW\ as well as almost cross-linearly-polarized excitation and collection beams (see SI). This allows us to further increase the amplitude of optical Rydberg transitions. 

Strikingly, the reflectance contrast amplitude $A_{n\mathrm{s}}$ of the Rydberg excitons undergoes periodic oscillations with \vtg, which arise due to the Landau-quantization of the graphene density of states. In general, at non-integer fillings the electrons are compressible and thereby efficiently screen the Rydberg excitons, which in turn quenches their optical response. Conversely, when the Fermi level lies in the gap between the LLs, the electronic state becomes incompressible, which results in a prominent enhancement of the amplitude of Rydberg transitions at subsequent integer filling factors. The $A_{n\mathrm{s}}$ maxima at $\nu=\pm2$ extend over a considerably broader \vtg\ range than those at other fillings. This is a consequence of non-uniform energy spacing of graphene LLs that are grouped into manifolds consisting of four LLs~\cite{DasSarma2011,Du2009,Bolotin2009}. The spin and valley degeneracy of LLs belonging to each manifold is lifted predominantly by many-body exchange interactions between the electrons~\cite{Young_Nature_2012}, but the resulting splitting still remains more than an order of magnitude smaller than the cyclotron gaps between subsequent manifolds. For $\nu=\pm2$, the LL spacing is $\Delta E_{\pm2}=\sqrt{2 e \hbar v_\mathrm{F}^2 B}\approx0.16\eV$ at $B=16\tesla$, where $\hbar$ denotes the reduced Planck constant, and $v_F$ is the Fermi velocity. To cross such a sizeable energy gap, we need to increase the gate voltage by $\Delta \vtg\approx0.4\volt$, which leads to the excessive width of $\nu=\pm2$ peaks. The analysis of the resulting displacement of other IQH features (see SI for details) allows us to extract $v_\mathrm{F}=(1.1\pm0.1)\times10^6\, \mathrm{m/s}$, which is consistent with previous reports for hBN-encapsulated graphene~\cite{Yu_PNAS_2013}.

Remarkably, the data in \figref{fig:Fig2}{\bf a} also feature another sequence of weaker enhancements of Rydberg exciton amplitude, hinting at the presence of additional incompressible states at fractional fillings. To get better insight into the nature of those states, we extract an effective measure for the 4s exciton amplitude defined as $A_\mathrm{4s}=\langle \Delta R/R_0 \rangle^\mathrm{peak}_\mathrm{4s} - \langle \Delta R/R_0 \rangle^\mathrm{dip}_\mathrm{4s}$, where the two terms represent the reflection contrast spectrally averaged over the peak and the dip around the 4s resonance (see SI). This procedure allows us to reduce the common mode noise, originating e.g. from the long-term fluctuations in our setup. The resulting \vtg-dependence of $A_\mathrm{4s}$ (\figref{fig:Fig2}{\bf b}) indeed exhibits pronounced peaks at voltages precisely corresponding to either $1/3$ or $2/3$ fillings of all subsequent LLs. This observation provides unequivocal evidence for the formation of FQH states in graphene. Interestingly, the FQH states at the first excited LL1 manifold ($2<|\nu|<6$) give rise to significantly stronger Rydberg exciton amplitude maxima as compared those at LL0 ($|\nu|<2$); this is unexpected as prior transport experiments have demonstrated that FQH states at LL0 are slightly more robust~\cite{Polshyn2018}. Temperature dependent measurements we discuss below indicate that the activation gaps of FQH states at LL0 and LL1 are also comparable in our sample, which suggests that striking difference in FQH visibility at $|\nu|<2$ and $|\nu|>2$ is an inherent property of the Rydberg exciton sensing scheme we employ.

Taking advantage of the fact that FQH signatures are particularly well-resolved at $2<\nu<3$, we examine this filling-factor range closer by averaging reflectance spectra acquired with longer integration time, which allows us to further improve the signal-to-noise ratio. As shown in Figs\,\ref{fig:Fig2}{\bf c,d}, this reveals the presence of even fainter $A_{4\mathrm{s}}$ maxima corresponding to FQH states with larger denominators of 5 and 7. We emphasize that the number of fractional fillings we observe are comparable to the highest odd denominators that have been previously reported for graphene using transport studies at similar magnetic fields~\cite{Polshyn2018,Zeng2019}, suggesting that Rydberg exciton spectroscopy yields a sensitivity that is similar to state-of-the-art transport techniques.

Having established optical access to FQH states in graphene, we investigate their robustness to changes in system parameters. \figref{fig:Fig3}{\bf a} presents the voltage-dependent amplitude of 4s resonance measured at the same $T=80\mkelvin$, but at different magnetic fields. As expected, when lowering the $B$-field, signatures of FQH states become less prominent and are no longer discernible at $B=6\tesla$. In parallel, owing to reduced LL degeneracy (being proportional to $B$), the voltage gaps between subsequent Rydberg amplitude maxima become narrower. To analyze this effect quantitatively, we fit the positions of the maxima with a set of dependencies obtained within the parallel-plate capacitor model (see SI), which forms a graphene LL fan diagram. This allows us to extract the basic parameters of the graphene monolayer: in addition to $v_\mathrm{F}$ introduced earlier, we find the $\vtg \approx 22\mvolt$ corresponding to graphene charge-neutrality as well as the voltage change $\Delta \vtg\approx62.5\mvolt / \tesla$ required to fill a LL at $B=1\tesla$. The latter value can be used to determine the out-of-plane hBN dielectric constant to be $\epsilon^\perp_\mathrm{hBN}=3.1\pm0.2$; this is consistent with both previously reported values~\cite{Xu2021,Smolenski2019} and that obtained based on our observation of optical Shubnikov-de Haas oscillations of the 1s exciton transition in the TMD monolayer~\cite{Smolenski2019} region of the same sample (see SI). We emphasize that this model is used for assignment of all IQH and FQH states in our work. Its perfect agreement with the data further corroborates our identification of the amplitude maxima as FQH states, and demonstrates in particular that fractal Hofstadter states, which could arise from a moir\'e superlattice between hBN and graphene, play no role in our device~\cite{Nature_Ponomarenko_2013,Dean2013,Yu_NatPhys_2011}. We further exclude any influence from Landau-quantization of the top and bottom graphite gates~\cite{Zhu2021} on the investigated effects by performing analogous \vtg-dependent reflectance measurements at various back gate voltages (see SI).

The visibility of FQH states is also reduced at elevated temperatures. \figref{fig:Fig3}{\bf b} shows a set of Rydberg optical responses acquired at fixed $B=16\tesla$, but for various device temperatures. The peaks associated with $n/3$ FQH states disappear around $T\approx4-6$~K for all investigated LLs at $|\nu|<5$, which corresponds to activation energy gaps a few times smaller than the highest gaps reported for the same FQH states in previous transport studies of graphene at comparable $B$-fields~\cite{Polshyn2018}. We also note that owing to the small but finite graphene light absorption, the signatures of FQH states are sensitive to the optical excitation power, becoming sizeably suppressed upon raising the power to few tens of nW (see SI). 

Our findings have been reproduced on a few different spots in the investigated sample region featuring 5ML-thick hBN spacer between the graphene and the TMD monolayer. Interestingly, we have not been able to detect FQH states in the 2ML-thick spacer region with similar acquisition time. This is most likely a consequence of the strongly suppressed amplitude of Rydberg transitions in this region even for integer fillings, which in turn reduces the sensitivity of our optical probing technique. Given that Rydberg resonances also become insensitive to the graphene compressibility for spacer thicknesses approaching their Bohr radii limited by the magnetic length, these findings illustrate that there is an optimal spacer thickness which maximizes the visibility of FQH states for any given $B$ (see SI). Finally, we emphasize that we have also observed analogous signatures of the formation of FQH states in Bernal-stacked bilayer graphene with a 5ML-thick spacer, which further confirms the applicability of our detection scheme for various two-dimensional materials.

The experiments we report in this work establish the TMD monolayer as a non-destructive optical sensor for the FQH states in an adjacent graphene layer. This remote sensing scheme based on Rydberg excitons demonstrates the feasibility of optical investigations of strong electronic correlations in atomically-thin materials that are otherwise optically inaccessible, e.g., indirect band-gap semiconductors or MATBG. In parallel, Rydberg exciton spectroscopy offers excellent spatial resolution determined by sub-micron diffraction limit for the optical spot size. Compared to conventional transport techniques, this allows to greatly reduce the influence of spatial inhomogeneities that are inherent to most van der Waals heterostructures. This may in turn pave the way for the observation of novel exotic correlated phases, such as fractional Chern insulators~\cite{Spanton_science_2018,Xie2021} in MATBG which have been demonstrated to be very sensitive to twist-angle variations occurring at \textapprox $\mu$m length scales~\cite{Uri2020}.

Remarkably, our approach also provides ultra-fast temporal resolution determined by the picosecond-long exciton lifetime. This brings up a unique opportunity to explore nonequilibrium dynamics of strongly correlated electronic phases that is beyond reach of any transport experiments. Another promising extension of our optical detection scheme would be to replace the sensing layer with a TMD bilayer structure hosting optical excitations that are coherent superpositions of intra- and inter-layer excitons~\cite{Shimazaki2020,Zhang_Nature_2021}. Due to the finite dipole moment of these hybrid excitons, we expect this scheme to allow for optical explorations of electronic correlations with drastically enhanced sensitivity.

\begin{acknowledgments}
We thank Puneet Murthy, Thibault Chervy, Xiaobo Lu and Titus Neupert for fruitful discussions. This work was supported by the Swiss National Science Foundation (SNSF) under Grant No. 200021-178909/1. K.W. and T.T. acknowledge support from the Elemental Strategy Initiative conducted by the MEXT, Japan (Grant Number JPMXP0112101001) and JSPS KAKENHI (Grant Numbers 19H05790 and JP20H00354).
\end{acknowledgments}

\end{document}



\title{Supplementary Information for \\ ``Optical sensing of fractional quantum Hall effect in graphene''} 

{\footnotesize 
\author{A. Popert}
\affiliation{\ETH}

\author{Y. Shimazaki}
\affiliation{\ETH}
\affiliation{\RIKEN}

\author{M. Kroner}
\affiliation{\ETH}

\author{K.~Watanabe}
\affiliation{\NIMSRCFM}

\author{T.~Taniguchi}
\affiliation{\NIMSICMN}

\author{A. Imamo\u{g}lu}
\affiliation{\ETH}

\author{T. Smole\'nski}
\affiliation{\ETH}

\maketitle

\tableofcontents
}

\newpage

\section{Sample and characterization}
\label{sec:sample}

\subsection{Device fabrication}

\begin{figure}[b!]
    \includegraphics[width=1\textwidth]{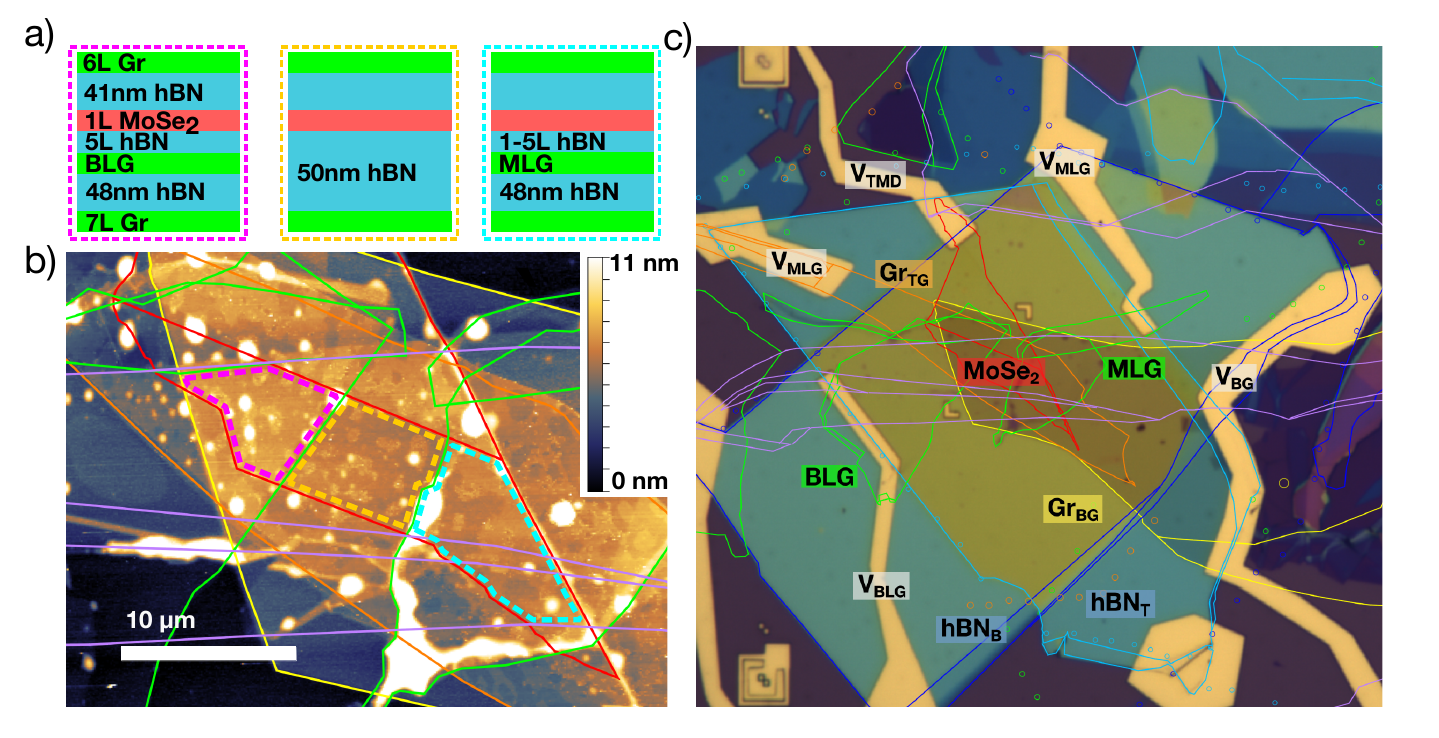}
    \caption{\label{fig:sample_geometry}{\bf Sample geometry.} {\bf a} Schematics of the layer structure of different sample regions. \MoSe\ flake is lined in red, graphene flakes in green, and hBN flakes in blue. {\bf b} AFM image of the sample with the different regions indicated by dashed lines. {\bf c} Optical microscope image of the sample overlaid with the stacking plan after the metal deposition. The labels indicate: monolayer graphene (MLG), bilayer graphene (BLG), graphite top and bottom gates (Gr$_\mathrm{TG}$, Gr$_\mathrm{BG}$), hBN top and bottom dielectrics (hBN$_\mathrm{T}$, hBN$_\mathrm{B}$), the voltage applied to the multilayer part of the TMD (V$_\mathrm{TMD}$), to the proximal graphene (V$_\mathrm{MLG}$), to the proximal bilayer graphene (V$_\mathrm{BLG}$), to the back and top gates (V$_\mathrm{BG}$, V$_\mathrm{TG}$). During all of the reported experiments V$_\mathrm{TMD}$, V$_\mathrm{MLG}$, V$_\mathrm{BLG}$ are set to 0\volt.}
\end{figure}

For assembly of our heterostructure, we use the dry stacking technique with a dome-shaped PDMS stamp~\cite{Lee2014a, Kim2016} coated with a sacrificial polycarbonate (PC) layer. To minimize contamination, the stacking is carried out at high temperatures (\textapprox 100$^\circ$-130$^\circ$~C)~\cite{Pizzocchero2016} in an Argon-filled glovebox. Moreover, we choose flakes that show no signatures of exfoliation-tape residues. The Cr:Au (5\nm\ : 45\nm) electrodes are patterned with standard electron-beam lithography. 

The sample structure is shown in Fig.\,\ref{fig:sample_geometry}. It consists of a \MoSe\ monolayer (ML) which is separated from a graphene by a thin hBN spacer. The whole structure is embedded in a \textapprox 41\nm\ top hBN flake and a \textapprox 48\nm\ bottom hBN flake. For gating, we use 6-monolayer-thick (6ML) graphite as a top gate and 7ML graphite as the bottom gate. We use graphite flakes for the gates because they provide a higher mechanical stability and tend to fold or break less than a ML graphene. Furthermore, thicker graphite is more efficient at screening environmental disorder.

\begin{figure}[b!]
    \includegraphics[width=0.6\textwidth]{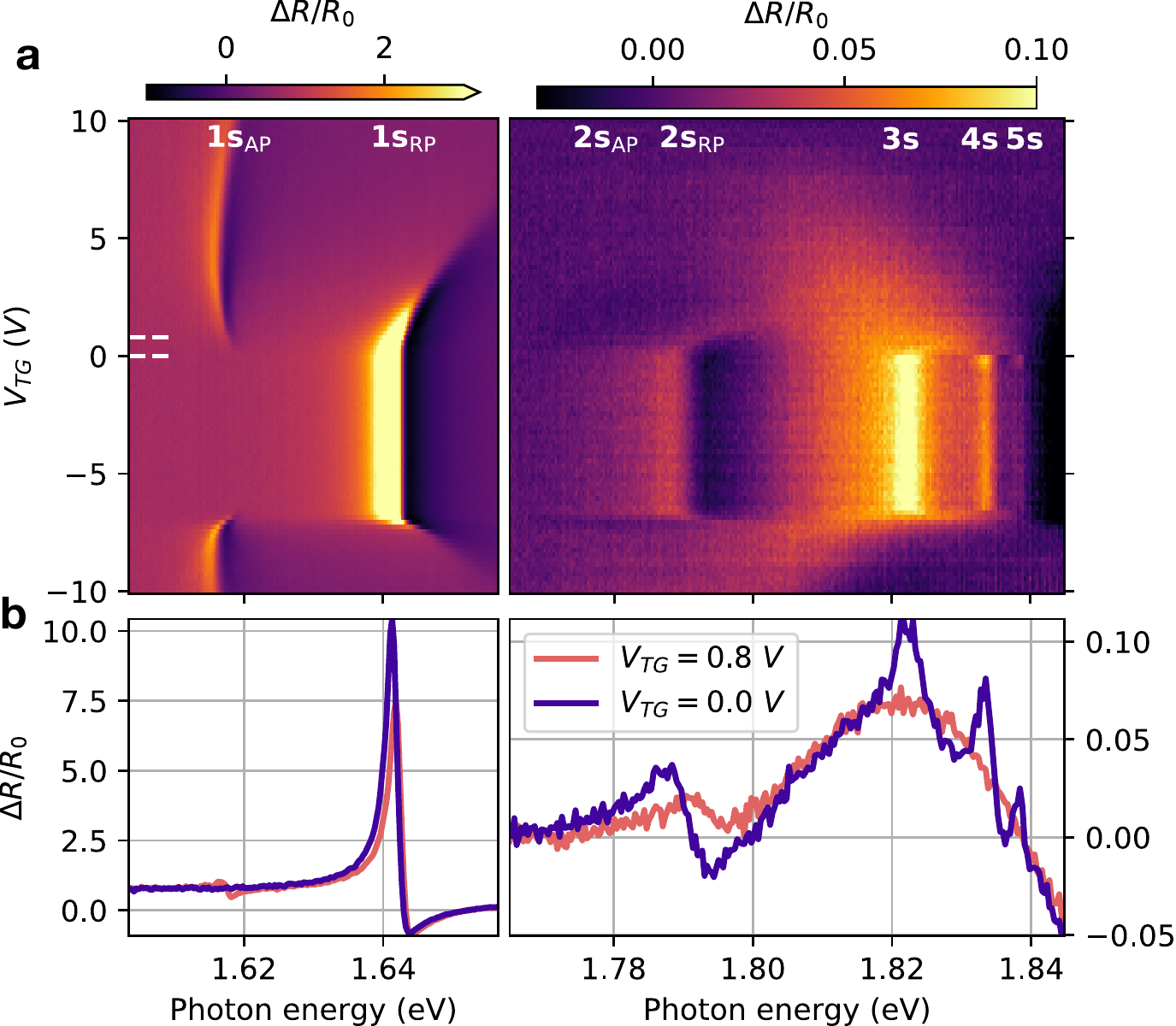}
    \caption{\label{fig:0T_control_region}{\bf Optical response of the control region of the investigated device.} (a) Top-gate-voltage evolution of the reflectance contrast spectra measured at $V_\mathrm{BG}=0$, $T=4$~K in the spectral ranges of the 1s exciton (left panel) and Rydberg excitons (right panel). The data was acquired in the control region of our device without proximal graphene layer. (b) Line cuts through the maps in (a) at two different voltages corresponding to charge neutrality ($V_\mathrm{TG}=0$) and slight electron doping ($V_\mathrm{TG}=0.8\volt$). In the latter case, the 3s to 5s excitons are completely quenched, leaving only the broad contribution from the B-exciton.}
\end{figure}

The sample hosts multiple regions (see Fig.\,\ref{fig:sample_geometry}\textbf{a}). A control region with \MoSe\ without proximal graphene is used for the reference purposes (see Fig.\,\ref{fig:0T_control_region}). In the region covered by graphene, there are steps in hBN thickness, allowing us to vary the distance between the graphene and the TMD simply by investigating different spots on the sample. Most of our measurements are carried out in a region with a 5ML-thick hBN spacer --- for simplicity we refer to it as the graphene-covered region. We also perform some measurements in the region with a 2ML-thick hBN spacer (see Sec.~\ref{sec:other_areas}). Finally, the device also hosts a region with a Bernal-stacked bilayer graphene (BLG) proximal to the TMD flake, separated by a 5ML-thick hBN spacer, which we refer to as the BLG-covered region. Having multiple regions allows us to explore various geometries within a single device without having to exchange the sample between the measurements. 

We estimate the approximate thicknesses of the flakes just after the exfoliation. To this end, we compare the reflection contrast between the flake and the SiO$_2$/Si substrate measured on each of the RGB channels of our CCD camera with pre-calibrated reference values~\cite{Golla2013, Smolenski2019}. The precise thicknesses of all flakes are later determined based on the atomic-force microscopy (AFM) of the assembled stack before patterning the gate electrodes. Before assembly of our heterostructure, we perform coupled-capacitor simulations~\cite{Barrera2017} to estimate the achievable doping levels for given hBN thicknesses. We particularly ensure that the MoSe$_2$ monolayer in the 5ML-thick hBN spacer region can be electron-doped before reaching the hBN breakdown electric field, which is between 0.15 and 0.25\volt/\nm~in our devices. We emphasize that thinner hBN gate dielectrics reduce the maximum achievable doping in TMD monolayers.

\subsection{Experimental setup}
\label{sec:setup}

Our measurements are performed in a dilution refrigerator immersed in a Helium bath cryostat equipped with a superconducting magnet, which allows us to apply external magnetic fields of up to 16\tesla\ in the direction perpendicular to the sample surface. The refrigerator is fitted with a 100~$\Omega$ resistor enabling us to heat the sample and to set the base temperature within the range between 80\mkelvin\ and 20\kelvin. Our dilution unit also has an optical access via a single-mode optical fiber. The light is coupled out from the fiber and focused onto the sample through a home-built cryogenic objective consisting of two lenses with NA = 0.15 (fiber side) and NA = 0.68 (sample side). This configuration permits to achieve an optical spot with a typical diameter of $\sim0.8\um$. The objective is mounted on cryogenic piezoelectric stepper actuators allowing us to precisely select the analyzed spot on the sample surface. 

The light reflected off the sample is transmitted through the same fiber. It is then separated from the exciting beam in a free-space optical setup that also enables us to control the polarization of both beams (see Refs~\cite{Knuppel2019,Smolenski2021} for more detailed description). Since a single-mode fiber acts as a strongly chromatic high-order waveplate, we use two different polarization settings for measuring 1s exciton and Rydberg excitons. In both cases, the polarization is adjusted based on the exciton/polaron resonances observed in the respective spectral range at high magnetic fields. Specifically, we set the polarization to be circular in order to maximize the contrast between Zeeman-split branches of the optical resonances. Note that in most of our measurements of the Rydberg states, the excitation and collection beams are nearly cross-linearly-polarized, as described in Sec.\,\ref{sec:data_analysis}. The acquired reflectance signal is spectrally resolved with 0.5\,m spectrometer and then recorded with liquid-nitrogen cooled CCD camera.

As a white light source for our reflection contrast measurements in the 1s spectral range, we use a narrow-band light emitting diode (LED) centered around 760\nm\ with 3-dB bandwidth of $\textapprox20\nm$. For the spectral range of the Rydberg excitons, we use either a halogen light source or a supercontinuum laser, which is filtered spectrally to minimize the overall power of light illuminating the sample.


\subsection{Basic characterization}
\label{sec:characterization}

\begin{figure}[b!]
    \includegraphics[width=0.8\textwidth]{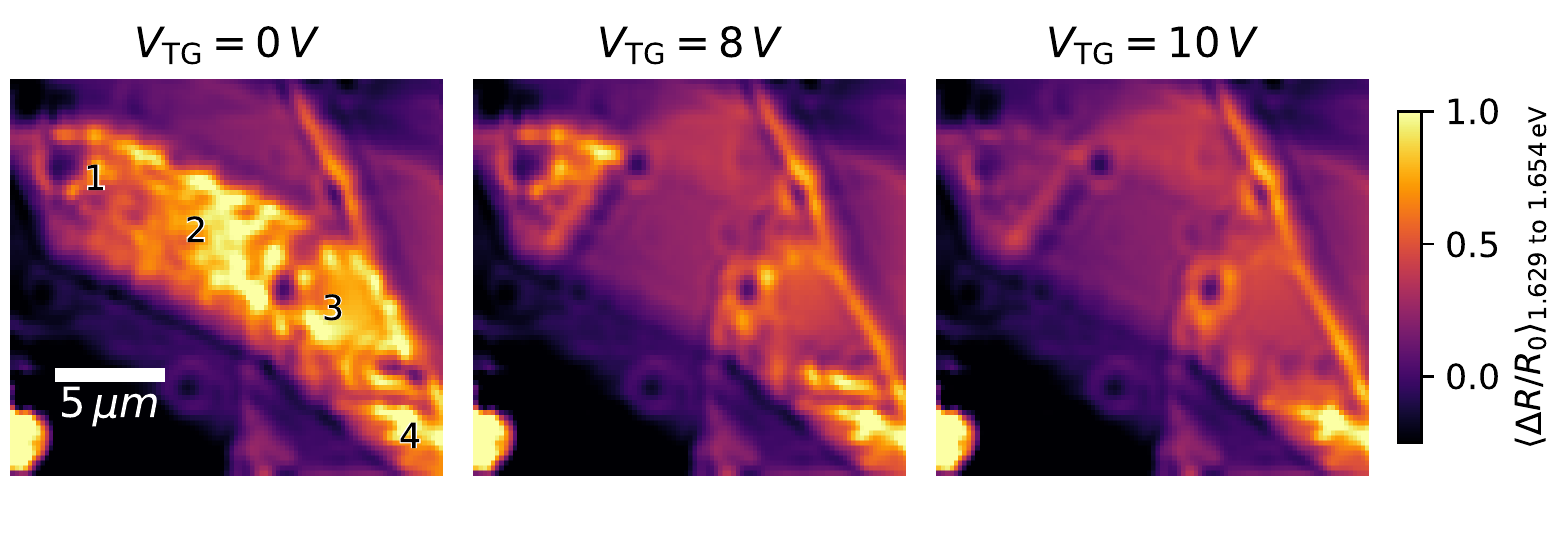}
    \caption{\label{fig:characterization_vtg_map}{\bf Spatial map of the reflectance contrast in the 1s exciton spectral range at different top gate voltages.} The panels show \DRR\ averaged over the 1s exciton spectral region for $\vbg=0\volt$. A strong reflectance signal from the exciton corresponds to charge neutrality. When the TMD monolayer in a given region is electron doped, the signal decreases. In the control region (labelled as ``2''), the 1s exciton is completely quenched when applying $\vtg=8\volt$. In the BLG-covered region (``1''), the 1s exciton only starts to be quenched at $\vtg=10\volt$, while the TMD monolayer in graphene-covered region with 2ML-thick hBN spacer region (``4'') remains charge neutral over the whole \vtg\ range. Finally, the TMD in the 5ML-thick hBN spacer region (``3'') is weakly doped at $\vtg=8\volt$ and becomes more strongly doped at $\vtg=10\volt$.}
\end{figure}

A typical top gate voltage evolution of the \MoSe\ reflection contrast spectrum in the control region (without proximal graphene) is shown in Fig.~\ref{fig:0T_control_region}. In the charge neutrality region ($-7\volt \leq \vtg \leq 0.4\volt$), the TMD is devoid of itinerant charge carriers and its reflectance spectrum exhibits a series of neutral exciton resonances: the 1s exciton transition is shown in the left panels, the Rydberg excitons from 2s to 5s are presented in the right panels. 

We note that 1s exciton in our device shows a large reflection contrast owing to small inhomogeneous broadening and a particular choice of hBN top and bottom layer thicknesses that suppresses the overall reflection coefficient of the multilayer stack, thereby enhancing the signal originating from the TMD flake~\cite{Fang_PRL_2019,Smolenski2019}. 

Upon electron doping ($\vtg > 0.4\volt$), the 1s exciton becomes dynamically screened by injected electrons~\cite{Sidler2017,Efimkin2017} and evolves into a repulsive Fermi-polaron (1s$_\mathrm{RP}$) that blueshifts and becomes progressively weaker. In parallel, its oscillator strength is transferred to the second, attractive polaron (1s$_\mathrm{AP}$) branch that appears at lower energies. For the 2s exciton, the attractive polaron transition is significantly broader and only weakly visible~\cite{Wagner2020,Liu_NatCommun_2021}. On the other hand, higher Rydberg states (3s to 5s) abruptly disappear upon electron doping, and the signatures of their respective attractive polaron branches are not discernable in our data. Note that these Rydberg transitions are overlaid on a background stemming from the \MoSe\ B-exciton~\cite{Goldstein2020}. This resonance is broad and much more weakly affected by doping than the Rydberg states. For this reason, it has a negligible effect on our data analysis (see Sec.~\ref{sec:data_analysis}). 

The different sample regions show stark differences in doping characteristics, as we illustrate in Fig~\ref{fig:characterization_vtg_map}. In the control region, we only need to apply a small \vtg\ to introduce electrons to the TMD monolayer. In the regions where there is a proximal graphene, electron doping of the TMD occurs at higher voltages owing to sizeable energy offset between the graphene and the TMD conduction bands~\cite{Wilson2017}. We emphasize that in each region the gate voltage corresponding to the onset of electron doping is well defined, uniform over the respective sample areas, and consistent with coupled capacitor simulations~\cite{Barrera2017}.

An important experimental challenge for reproducibility of our measurements is a precise alignment of the position of the optical spot after changing the system parameters (e.g., elevating the temperature or taking the reference spectrum out of the MoSe$_2$ flake). Since our dilution refrigerator does not have imaging access, we use the \DRR\ spectrum itself to navigate over the sample surface. Specifically, we utilize the onset of electron doping to determine the sample region of the currently investigated spot. For reproducible position alignment within a given region, we take advantage of the fact that the \DRR\ spectral profile of the 1s exciton shows large spatial variations in energy, amplitude and linewidth. By matching the \DRR\ spectrum with a reference one, we can therefore precisely realign the position of our optical spot.

\subsection{Electron density calibration and LL filling factor assignment}
\label{sec:density_calib}

We calibrate the electron densities in our device using two different methods. This enables us to precisely relate the applied gate voltages to the corresponding LL filling factor. Furthermore, this is a verification that the proximal \MoSe\ layer does not affect the graphene filling factor.

In the first method, we analyze the LL fan chart of the proximal graphene layer obtained using our analysis of the optical response of Rydberg excitons. To this end, we first extract the top-gate voltages $\vtg (\nu,B)$ corresponding to signatures of IQH and FQH states in graphene at various filling factors $\nu$ and magnetic fields $B$. We obtain these voltages by fitting the positions of the 4s exciton amplitude maxima in the data from Fig.~\,3{\bf a} with phenomenological Gaussian profiles. These voltages are then fitted with a set of dependencies that form the graphene LL fan chart, given by
\begin{align}
\label{eqn:ll_fan_chart}
    \vtg(\nu)=V_0 + \nu \Delta V_\mathrm{LL} B + \frac{E_\mathrm{LL}(\nu)}{e} \left(1+\frac{t_T^*}{t_B}\right),
\end{align}
where $E_\mathrm{LL}(\nu)$ is the energy of the corresponding LL. In our analysis we neglect a small, interaction-induced energy splitting between the graphene LLs belonging to each manifold and consider exclusively the large cyclotron gaps between subsequent manifolds. For the first two LL manifolds ($ |\nu|\leq 6$), the energies are given by
\begin{align}
    E_\mathrm{LL}(\nu) = \left\{
    \begin{array}{cc}
    0 & \mathrm{for}\ |\nu|\le2,\\
    \mathrm{sgn}(\nu)\sqrt{2 e \hbar v_\mathrm{F}^2 B} & \mathrm{for}\ 2 < |\nu| \leq 6.
    \end{array} \right.
\end{align}
The graphene LL fan chart has three fitting parameters: $V_0=(22\pm2)\mvolt$ is the voltage corresponding to graphene charge-neutrality, $v_\mathrm{F}=(1.1\pm0.1)\times10^6\ \mathrm{m/s}$ is the Fermi velocity, and $\Delta V_\mathrm{LL}=(62.5\pm0.2)\mvolt / \tesla$ is the voltage change required to fill a graphene LL at $B=1\tesla$. The values of $t_B\approx48\nm$ and $t_T^*\approx45\nm$ represent, respectively, the thickness of the bottom hBN layer and an effective thickness of the hBN layer separating the graphene and from the top gate. $t_T^*$ is given by the sum of the thicknesses of the top hBN layer (41\nm), 5ML hBN spacer ($1.7\nm$) and a contribution stemming from the presence of the TMD monolayer $d_\mathrm{TMD}\epsilon^\perp_\mathrm{TMD}/\epsilon^\perp_\mathrm{hBN}\approx2\nm$, where $d_\mathrm{TMD}$ denotes the MoSe$_2$ layer thickness of $0.7\nm$, while $\epsilon^\perp_\mathrm{TMD}\approx7.2$~\cite{Laturia_2DMA_2018} is its out-of-plane dielectric constant in the monolayer limit. The rightmost term in Eq.\,(\ref{eqn:ll_fan_chart}) contains the scaling factor $\Delta\vtg/\Delta E_F = \left(1+t_T^*/t_B\right)/e$ between the change of the top gate voltage $\Delta\vtg$ and a corresponding Fermi level shift $\Delta E_F$ under an assumption that the graphene is in a gapped state. We emphasize that the above expression was used to compute the voltages corresponding to integer and fractional filling factors in all figures in the main text and the supplementary information.

Given the LL degeneracy $eB/h$, the extracted value of $\Delta V_\mathrm{LL}$ can directly be related to geometrical capacitance between the graphene and the top gate electrode $C_\mathrm{geom}/e=(e/h)/\Delta V_\mathrm{LL}$. Within the frame of a parallel-plate capacitor model, this quantity can be in turn expressed as $C_\mathrm{geom}=\epsilon_0\epsilon_\mathrm{hBN}^\perp/t_T^*$, where $\epsilon_0$ denotes vacuum permittivity. Combining these two expressions, we obtain the value of out-of-plane hBN dielectric constant $\epsilon_\mathrm{hBN}^\perp=e^2t_T^*/h\epsilon_0\Delta V_\mathrm{LL}=3.1\pm0.2$, which is consistent with previous reports~\cite{Xu2021,Smolenski2019}. 

Note that when fitting the LL fan chart, we have not included the voltages corresponding to $\nu=\pm2$. These voltages are difficult to extract based on out data, since the corresponding Rydberg amplitude maxima form broad plateaus (see Fig. 3{\bf a} in the main text). However, we can use the width of these plateaus as an independent measure to extract the $v_F$. With the voltage broadening $\Delta V_{|\nu|=2}(B)=\sqrt{2\hbar e v_\mathrm{F}^2 B} (t_B+t_T^*)/et_B$ we therefore obtain $v_\mathrm{F}=(1.3\pm0.1)\times10^6\ \mathrm{m/s}$ which is similar to the value obtained in the first method. 

We have independently confirmed the above relation between \vtg\ and the electron density $n_e=\nu eB/h$ by measuring Shubnikov-de Haas oscillations of the 1s exciton transition at high magnetic field $B=16\tesla$ in the control region of our device. As demonstrated in our previous study~\cite{Smolenski2019}, the energy of the exciton resonance under the influence of such a high field exhibits prominent cusp-like shifts each time the system enters an IQH state. By tracing the voltages corresponding to those cusps, we can independently determine the voltage change $\Delta V^\mathrm{TMD}_\mathrm{LL}=(61\pm 5)\mvolt$ needed to fill a single LL, this time in the TMD monolayer. Given that in the investigated regime the electrons are fully spin- and valley-polarized, this allows us to extract the geometrical capacitance $C_\mathrm{geom}=e^2B/h\Delta V^\mathrm{TMD}_\mathrm{LL}$, which is now given by $C_\mathrm{geom}=\epsilon_0\epsilon_\mathrm{hBN}^\perp/t_T$ (note that since the electrons are now injected to the TMD monolayer, the relevant hBN thickness is the one of the top hBN flake $t_T\approx41\nm$). On this basis we again obtain $\epsilon_\mathrm{hBN}^\perp = 3.2\pm 0.3$, which is in good agreement with the value determined based on our investigations of graphene LLs.

We emphasize that in both of the above introduced methods of electron density calibration, we have assumed that there are no chargeable defects in the hBN or in the \MoSe, and that optical doping is negligible. The agreement of the extracted values of $v_\mathrm{F}$ and $\epsilon_\mathrm{hBN}^\perp$ with the previous reports shows that these two assumptions hold true in our device.

\section{Data analysis procedure}
\label{sec:data_analysis}

\subsection{Normalization of reflectance spectra}

In all of the reported experiments, we analyze the reflectance contrast $\Delta R/R_0$ of our sample, which is defined as
\begin{align}
    \frac{ \Delta R}{R_0} = \frac{R - R_0}{R_0},
\end{align}
where $R$ is the bare white-light reflection of the investigated spot, while $R_0$ represents the background reflection spectrum (corrected for the dark counts of our CCD camera). In general, $R_0$ should be selected such that it represents the reflection of the multilayer stack encapsulating the TMD monolayer, but is devoid of any spectral features arising from the TMD flake itself.

\begin{figure}[b!]
    \includegraphics[width=0.55\textwidth]{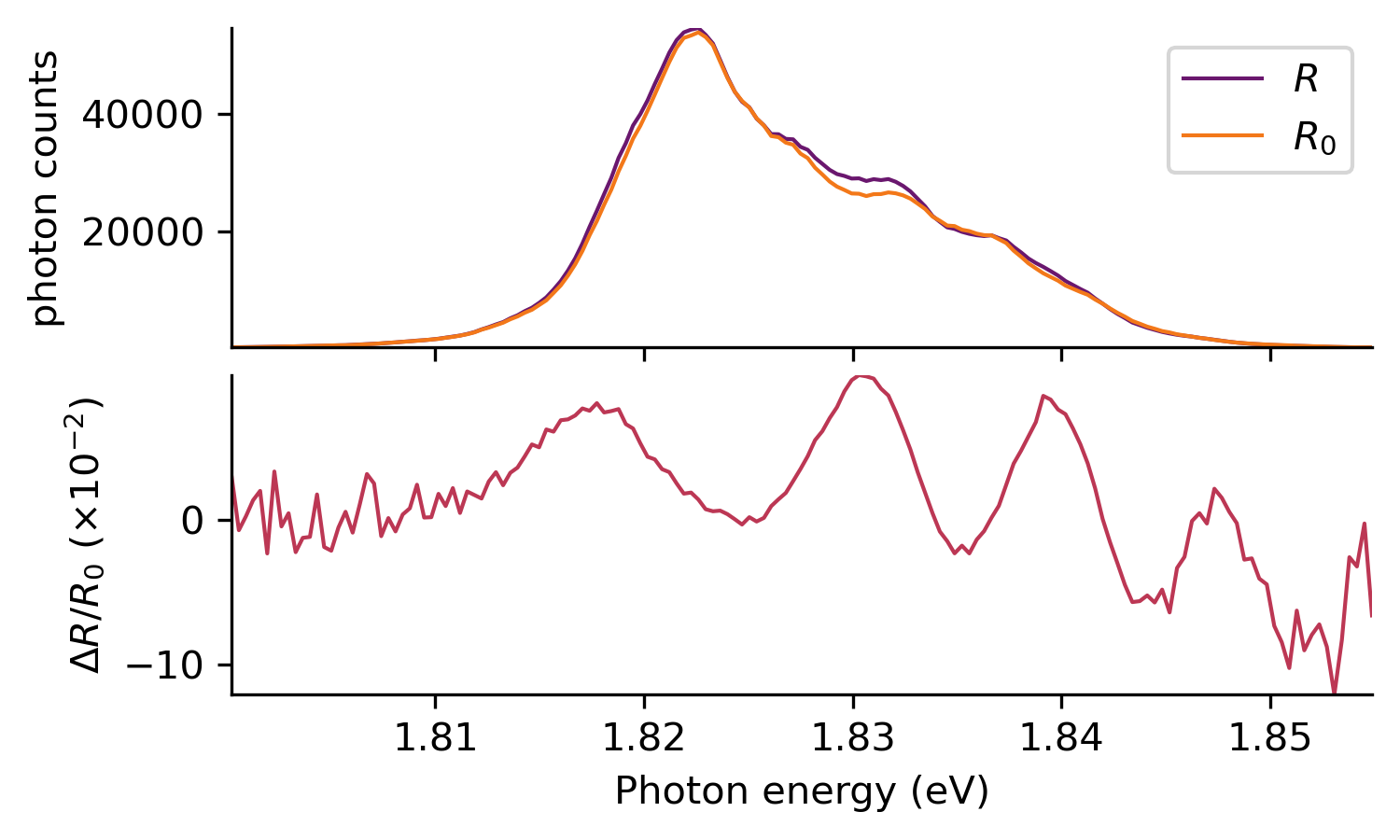}
    \caption{{\bf Normalization of reflectance spectra in the Rydberg spectral region at $T=80\mkelvin$ and $B=16\tesla$.} Upper panel: Two reflection spectra acquired in the graphene-covered region at the same spot, but for different top gate voltages: the reflectance signal $R$ with Rydberg excitons is taken at small $\vtg=-2.1\volt$, while the background spectrum $R_0$ is obtained at high $\vtg=9.5\volt$. Lower panel: Resulting reflection contrast spectrum. Note that the noise increases at the edges of the explored spectral window where the excitation white-light is filtered by a bandpass filter.}
    \label{fig:data_analysis_refcontrast_example}
\end{figure}

In our study, we use two different methods to obtain $R_0$ depending on the analyzed exciton species. For the 1s spectral region, the background spectrum is simply acquired at a different spot on the sample. This spot shows the same layer structure (i.e., hBN and graphene flakes) but without the TMD flake. This method is hardly applicable for the spectroscopy of Rydberg excitons, which exhibit much weaker \DRR\ making it difficult to reproducibly find the same spot on the sample based on the \DRR\ signal itself. For this reason, in our investigations of Rydberg excitons we pursue a different approach, in which we obtain a reference $R_0$ spectrum at the same spot as $R$, but for high $V_\mathrm{TG}>9\volt$. At this high voltage, the electrons are injected to the TMD monolayer even in the presence of a proximal graphene layer. This in turn quenches the optical response of the Rydberg excitons, thereby allowing to use the corresponding reflectance spectrum as $R_0$ for normalization purposes. Fig.\,\ref{fig:data_analysis_refcontrast_example} shows an example \DRR\ spectrum measured in the Rydberg spectral region with the corresponding $R$ and $R_0$ spectra plotted separately.

\begin{figure}[b!]
    \includegraphics[width=0.8\textwidth]{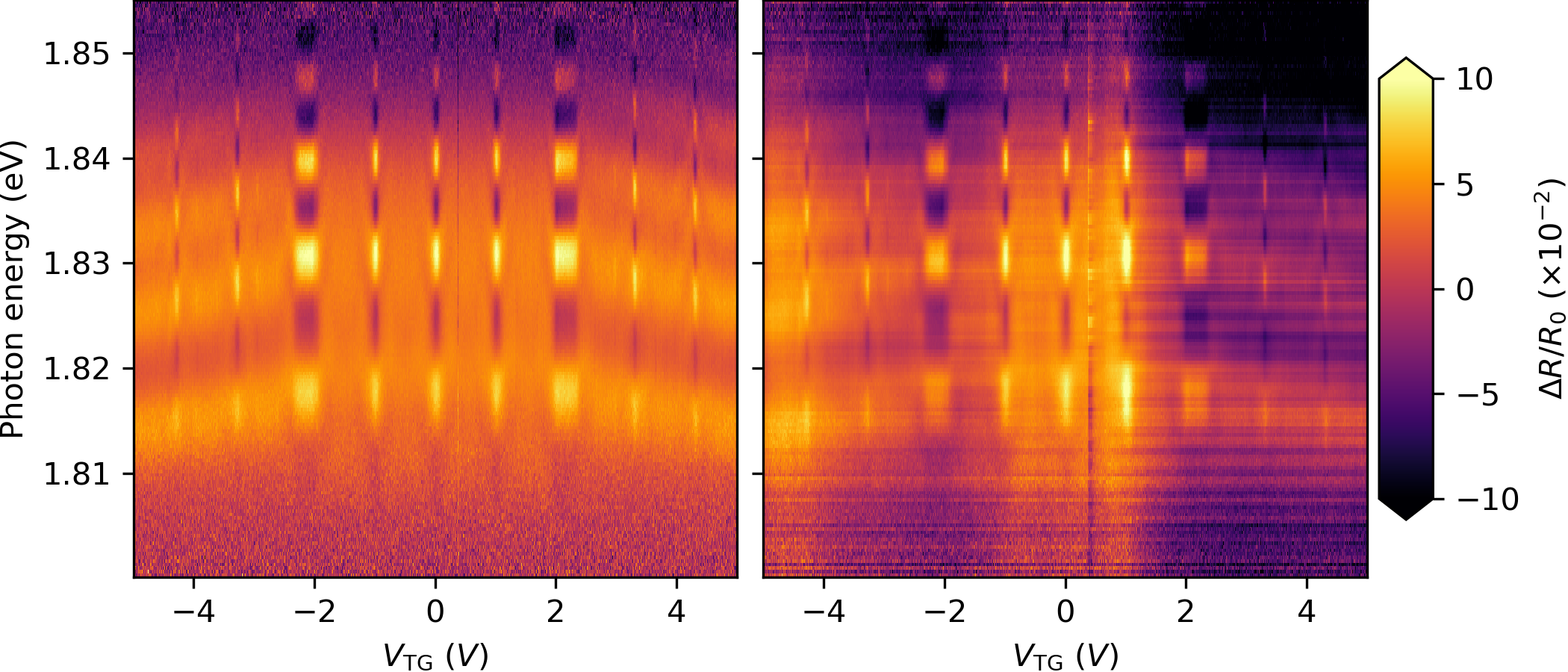}
    \caption{\label{fig:data_analysis_demonstrating_fringes}{\bf Effect of periodic background subtraction.} Left panel: Top-gate-voltage evolution of reflectance contrast spectra obtained at $T=80\mkelvin$, $B=16\tesla$ in the Rydberg spectral range using reference spectra taken periodically (approximately every minute) at $\vtg=9.5\volt$. Right panel: Same dataset, but with the reflection contrast signal determined using fixed background obtained before this \textapprox 48\,h-long measurement. Interference fringes stemming from etaloning-related effects in our optical setup appear as as horizontal lines that get more prominent over the course of the experiment (i.e., for larger gate voltages). This comparison demonstrates the importance of periodic background subtraction for obtaining low-noise reflectance contrast spectra in our experiments.}
\end{figure}

Since this normalization procedure exploits a reference spectrum taken at a high $\vtg$, it is not appropriate for the 1s exciton which remains bound even at the highest experimentally accessible gate voltages. This is shown in Fig.\,\ref{fig:0T_control_region}(a), where the 1s attractive polaron transition is still visible even at $\vtg=10\volt$. In principle, the same holds for the Fermi-polaron transition associated with the B-exciton, which spectrally overlaps with the Rydberg excitons and thereby might affect the $R_0$ spectrum taken at high $\vtg$. However, this effect is of negligible importance under our experimental conditions for two distinct reasons. First, the B-exciton transition is considerably broader than the Rydberg excitons. Second, in contrast to Rydberg exciton states, the B-exciton-polaron resonances are only weakly affected by electron doping, which in turn strongly reduces their influence on the $R_0$ spectrum obtained at small or moderate electron densities. This is illustrated in Fig.\,\ref{fig:0T_control_region}(b) showing \DRR\ spectra of Rydberg excitons in the control region using a reference spectrum $R_0$ measured at high TMD doping density $n_e\approx4\times10^{12}\ \mathrm{cm}^{-2}$ (for $\vtg=10\volt$). This choice of $R_0$ gives rise to a strong contribution from the B-exciton that appears as a broad shoulder in $\Delta R/R_0$ spectra at both $\vtg=0\volt$ and $\vtg=0.8\volt$. Conversely, if we used the spectrum at $\vtg=0.8\volt$ as $R_0$ (corresponding to weak electron doping $n_e\approx2\times10^{11}\ \mathrm{cm}^{-2}$), then the 3s--5s spectral range would be devoid of any contribution from the B-exciton. For our measurements in the main text, we utilize $R_0$ at moderate TMD doping levels $n_e \approx 1.5 \times 10^{12}\ \mathrm{cm}^{-2}$ corresponding to $\vtg=9.5\volt$ in the graphene-covered region. As a consequence, a weak broad residue of the B-exciton is still visible (as seen, e.g., in Fig.~2), but its influence on our analysis is negligible.

The main advantage of determining $R_0$ by changing the gate voltage is that this approach does not require to move the spot on the sample throughout the experiment, which provides an excellent long-term stability. This is particularly important for our measurements, which require obtaining $R_0$ periodically in order to suppress the interference fringes that emerge in our optical setup in a matter of minutes due to etaloning-related effects at wavelengths below 700~nm. For this reason, in all experiments on Rydberg excitons reported in the main text, we obtain a new $R_0$ spectrum roughly every minute. This allows us to greatly reduce interference effects, as seen in Fig.\,\ref{fig:data_analysis_demonstrating_fringes} presenting the comparison of the same $\vtg$-evolution of $\Delta R/R_0$ obtained either with periodically acquired background spectra or with the use of the same $R_0$ taken before the \textapprox 48\,h measurement.

\subsection{Polarization settings in FQH measurements}
\label{sec:pol_settings}

\begin{figure}[t!]
    \includegraphics[width=0.55\textwidth]{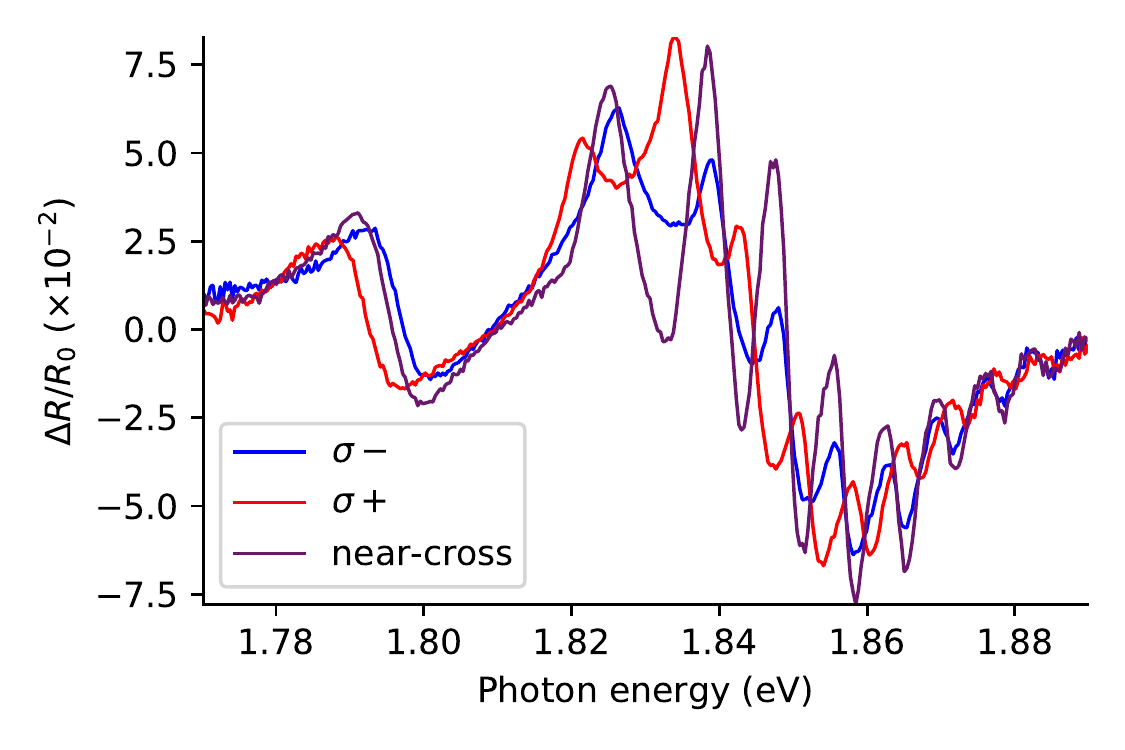}
    \caption{{\bf Increased signal-to-noise ratio of Rydberg amplitude for nearly cross-linearly-polarized excitation and collection beams.} Reflection contrast spectra obtained at $B=16\tesla$ and $T=80$~mK with the Fermi level lying in a graphene LL gap for different polarization settings. The two spectra are acquired with co-circular excitation/detection polarizations (either $\sigma^+$ and $\sigma^-$), while the third spectrum is taken with nearly cross-linearly-polarized excitation and collection beams. In circular polarization the 4s exciton $\Delta R/R_0$ amplitude yields $\textapprox 4 \%$, which gets enhanced by a factor $\textapprox 2$ for nearly cross-polarized settings.\label{fig:visibility_comparison_polarizations}}
\end{figure}

In all our FQH measurements, we use polarization settings in which the excitation and collection beams are partially cross-linearly-polarized, but still far from being fully polarization-suppressed to avoid strong reduction of the overall reflection signal. These polarization settings allow us to sizeably increase the reflectance contrast amplitude of the Rydberg exciton transitions compared to standard circular polarization basis, as shown in Fig.\,\ref{fig:visibility_comparison_polarizations}. The observed enhancement arises partially due to constructive interference between cross-circularly-polarized Zeeman-split branches of different Rydberg states. This is possible in the \MoSe\ monolayer at high magnetic fields, since the exciton valley-Zeeman splitting is nearly two times smaller than the energy splitting between subsequent Rydberg states, particularly for the states with moderate principal quantum numbers $3\lesssim n\lesssim 7$. Consequently, in this $n$-range, the $\sigma^+$-polarized branch of $n$s exciton appears almost precisely in the center between the energies of $\sigma^-$-polarized branches of $n-1$ and $n$ states. Owing to the dispersive lineshape of Rydberg resonances in our device, this leads to a situation in which, e.g., the peak of the 4s exciton in $\sigma^+$ polarization is almost perfectly aligned with the dip of the 3s transition $\sigma^-$ polarization. As such, by properly selecting the phase between the two components in partially cross-linearly-polarized settings, we can markedly increase the overall signal.

Note that owing to the contribution of the resonant fluorescence of Rydberg excitons, their visibility would increase even further for perfect cross-polarized settings, at the expense of a lower overall optical signal. Our polarization settings thus constitute a compromise allowing us to obtain a maximum signal-to-noise ratio for Rydberg excitons while keeping the excitation power sufficiently low.

\subsection{Extracting the reflectance contrast amplitudes $A_{n\mathrm{s}}$ of Rydberg exciton transitions}
\label{subseq:amplitude}

\begin{figure}[t!]
    \includegraphics[width=0.9\textwidth]{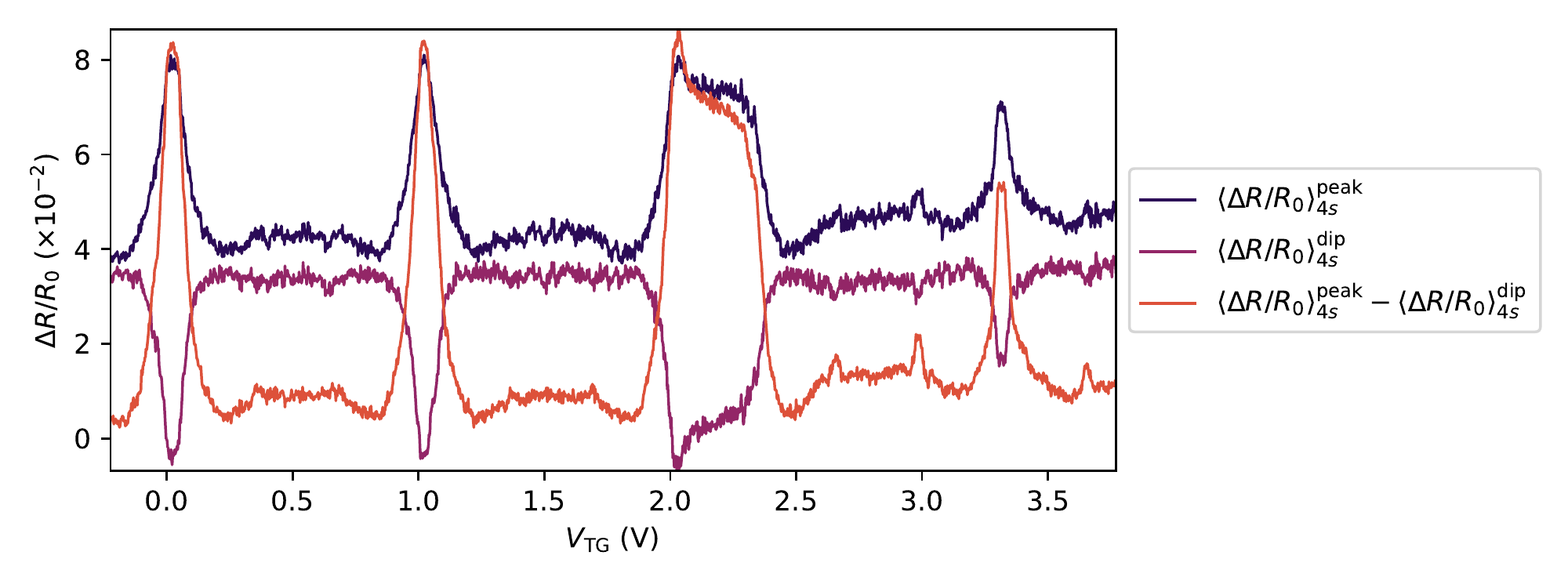}
    \caption{\label{fig:data_analysis_subtraction}{\bf Determination of effective amplitude of the 4s Rydberg excitons.} Gate-voltage evolution of the effective amplitude $A_{4\mathrm{s}}=\langle \Delta R/R_0 \rangle^\mathrm{peak}_{4\mathrm{s}} - \langle \Delta R/R_0 \rangle^\mathrm{dip}_{4\mathrm{s}}$ of the 4s exciton obtained based on the data from Fig. 2 in the main text. The two additional curves show separately the reflectance contrast signal averaged spectrally over the peak $\langle \Delta R/R_0 \rangle^\mathrm{peak}_{4\mathrm{s}}$ and dip $\langle \Delta R/R_0 \rangle^\mathrm{dip}_{4\mathrm{s}}$ component of the 4s resonance. The amplitude $A_{4\mathrm{s}}$ shows reduced noise and enhanced signatures of IQH and FQH states.}
\end{figure}

In our measurements, we use the reflectance contrast amplitudes $A_{n\mathrm{s}}$ of the Rydberg exciton optical transitions as a proxy for the compressibility of graphene electrons. In principle, these amplitudes could be extracted by fitting the measured spectra with multiple dispersive Lorentzian spectral profiles, as has been done in our previous studies of 1s excitons~\cite{Smolenski2019}. We exploit this procedure in our analysis of the zero-field data from Figs~1{\bf c} in the main text. However, in case of the data taken at $B>0$ we pursue a simpler approach, in which the amplitudes of Rydberg resonances are obtained by averaging the reflectance contrast signal around the transition energies. This is justified since at non-zero field we are interested exclusively in the amplitude of the optical resonances and not in their energies. In this procedure we take advantage of the fact that the Rydberg transitions in our device feature dispersive lineshapes, consisting of a peak and a dip components (as seen in Fig.\,\ref{fig:visibility_comparison_polarizations}). At each gate voltage, we manually select two, equally-wide spectral ranges around the peak and dip of a given Rydberg $n$s resonance (as schematically marked, e.g., in Fig.~2 in the main text). Then we average \DRR\ over these spectral ranges, which allows us finally to determine the effective Rydberg amplitude as a difference between the two values $A_{n\mathrm{s}}=\langle \Delta R/R_0 \rangle^\mathrm{peak}_{n\mathrm{s}} - \langle \Delta R/R_0 \rangle^\mathrm{dip}_{n\mathrm{s}}$. The main advantage of subtracting the peak and dip contributions is that this procedure eliminates common mode noise originating from long-term fluctuations of our setup, which may stem, e.g., from the temporal variations of the power of the light source. \figref{fig:data_analysis_subtraction} shows an example gate-voltage evolution of the amplitude $A_\mathrm{4s}$ of the 4s exciton obtained based on the dataset from Fig. 2 in the main text, together with the corresponding peak and dip contributions. The resulting $A_\mathrm{4s}$ shows more pronounced FQH signatures and its noise is close to the shot-noise limit.

\section{Characterization of the Rydberg excitonic sensing scheme}
\label{sec:technique}

\subsection{Sensitivities of different Rydberg excitons to IQH and FQH states in graphene}

\begin{figure}[t!]
    \includegraphics[width=0.8\textwidth]{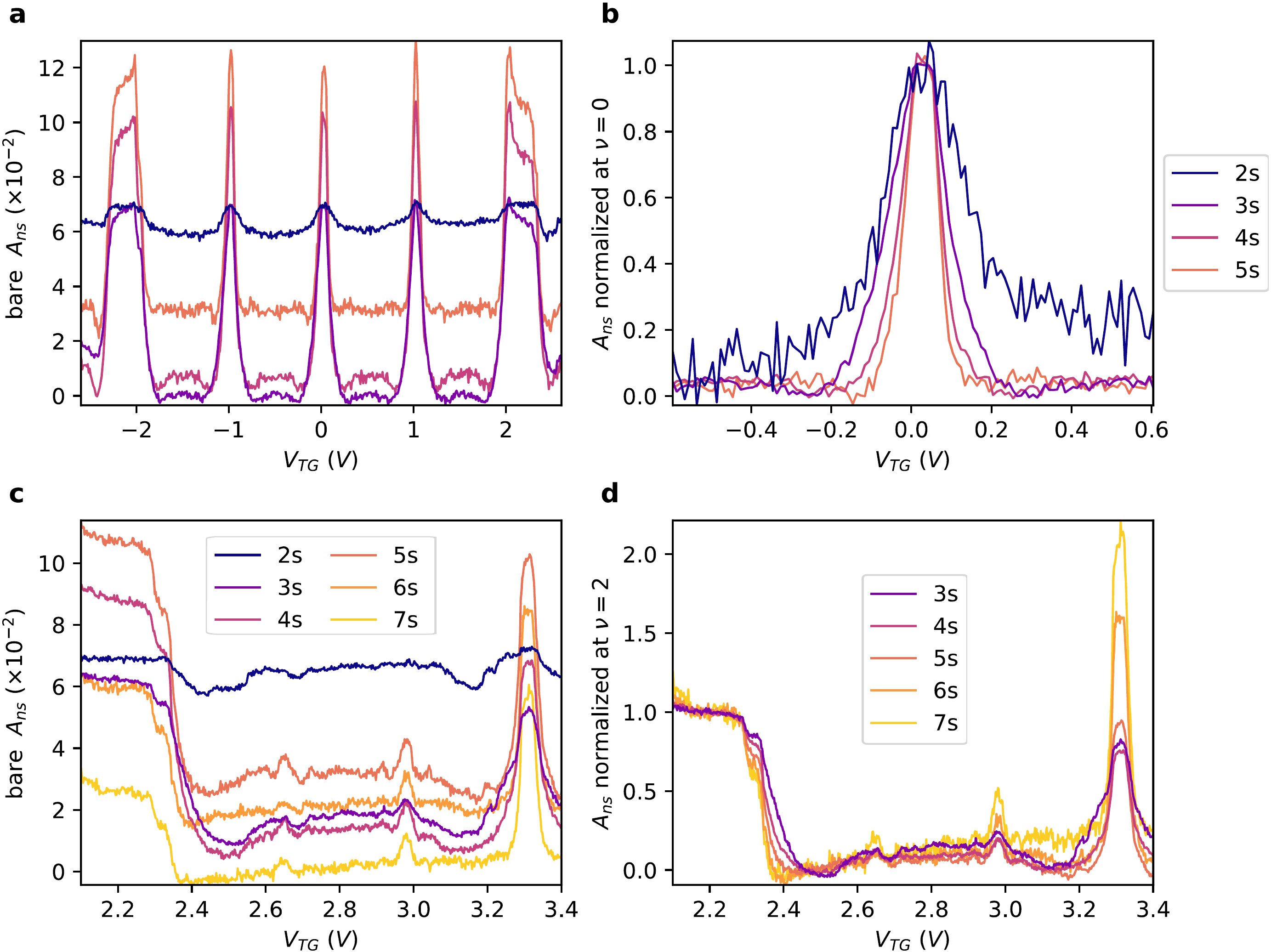}
    \caption{\label{fig:technique_quenching_behaviour}{\bf Sensitivity of different excitons to the formation of quantum Hall states.} Amplitudes of different Rydberg excitons $A_{n\mathrm{s}}$ extracted from datasets acquired in the 5ML-thick-spacer region at $B=16\tesla$, $T=80\mkelvin$ similar to Fig.\,2 from the main text, but for a white-light excitation in a broader, 50-nm-wide spectral range. {\bf a} $A_{n\mathrm{s}}$ as a function of \vtg\ in $-2 \leq \nu \leq 2$ for 2s to 5s Rydberg excitons. {\bf b} Same data plotted in a narrower voltage range around $\nu=0$ and normalized with respect to the height of $A_{n\mathrm{s}}$ maxima at $\nu=0$. {\bf c} \vtg-dependent $A_{n\mathrm{s}}$ for $2 \leq \nu \leq 3$ and different Rydberg excitons. {\bf d} Same as {\bf c} but normalized with respect to the height of $A_{n\mathrm{s}}$ maxima at the $\nu=2$.}
\end{figure}

Different Rydberg excitons exhibit different sensitivities to changes in the electron compressibility of the proximal graphene layer. For the choice of excitonic species in our FQH probing scheme at high magnetic fields, we need to consider two competing factors: the oscillator strength, and how strongly $A_{n\mathrm{s}}$ responds to changes in compressibility. 

\figref{fig:technique_quenching_behaviour}{\bf a},{\bf c} show the voltage-dependent $A_{n\mathrm{s}}$ determined at $B=16\tesla$, $T=80\mkelvin$ for various Rydberg states and two different filling factor ranges. We find that the 2s exciton is only weakly affected by the formation of IQH states, and therefore shows excessive noise compared to other exciton species when their response is normalized with respect to their $A_{n\mathrm{s}}$ maxima at $\nu=0$ (see \figref{fig:technique_quenching_behaviour}{\bf b}). Moreover, the amplitude $A_{2\mathrm{s}}$ of the 2s exciton does not display any observable signatures of FQH states, rendering it a poor probe of the compressibility. The 3s exciton does show FQH features, but they are still relatively weak. In general, these features tend to gain in visibility for Rydberg states with higher principle quantum number $n$ (see \figref{fig:technique_quenching_behaviour}{\bf d}), since such states are more sensitive to the FQH compressibility gaps owing to their smaller binding energies. These observations are contrasted by the fact that, for large $n$, the oscillator strength of Rydberg excitons becomes sizeably suppressed (see also \figref{fig:visibility_comparison_polarizations}). This results in a decrease of $A_{n\mathrm{s}}$, in turn reducing the overall sensitivity of high Rydberg excitons with $n>5$. 
These considerations yield the 4s and 5s excitons as the most sensitive probes of the compressibility of graphene electrons at $B=16$~T, which is the underlying reason for our choice of the 4s state in all of our FQH experiments reported in the main text. We emphasize that the optimal choice of Rydberg exciton species as compressibility probes generally depends on the magnetic field and on the thickness of the hBN spacer layer. Consequently, when designing our device we exploited Quantum Electrostatic Heterostructure (QEH) model~\cite{Gjerding2020} to estimate the optimal spacer thickness for our experimental conditions.

Finally, we note that quenching of $A_{n\mathrm{s}}$ when $E_F$ enters a graphene LL in general occurs more abruptly for higher Rydberg excitons. This can be seen in the normalized $A_{n\mathrm{s}}$ shown in \figref{fig:technique_quenching_behaviour}{\bf b}. When the LL starts to get filled, the graphene compressibility increases, and $A_{n\mathrm{s}}$ decreases. This decrease is faster for larger $n$, which manifests itself in reduced broadening of $A_{n\mathrm{s}}$ maximum for higher Rydberg states. This is consistent with their higher sensitivity to the changes in graphene compressibility.

\subsection{Influence of excitation power on FQH signatures}

\begin{figure}[t!]
    \includegraphics[width=0.45\textwidth]{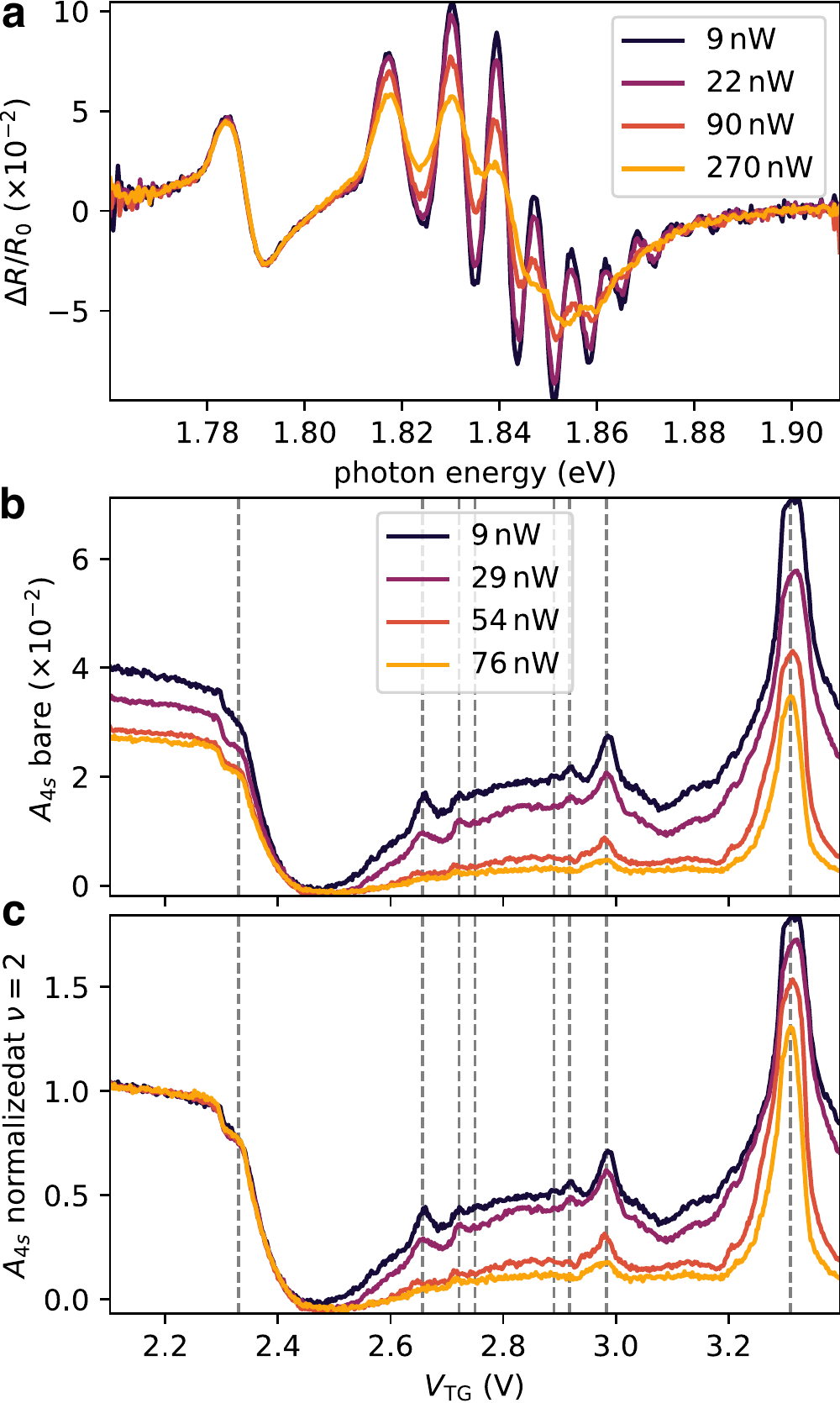}
    \caption{\label{fig:technique_power_dep}{\bf Influence of excitation power on the optical response of Rydberg excitons.} {\bf a}~Reflection contrast spectra of the Rydberg excitons acquired at various white-light excitation powers in the graphene-covered region at $B=16\tesla$, $T=80\mkelvin$, and $\nu=0$. Our optical spot size is $\textapprox 0.8\, \mathrm{\mu m}$. The Rydberg states with higher principal quantum numbers get quenched for powers exceeding 50\nW. {\bf b}~Effective amplitude of the 4s exciton measured at the same conditions as a function of \vtg\ for different excitation powers. {\bf c} The same curves as in ({\bf b}), but normalized with respect to the height of amplitude maximum at $\nu=2$ to compensate for power-induced reduction of the 4s exciton amplitude. Gray dashed lines in both panels indicate the positions of discernible FQH states. The signatures of all FQH start to quench already at 29\nW\ and become almost completely indiscernible at 76\nW. In addition, the states with higher denominators tend to disappear at lower powers.}
\end{figure}

As noted in the main text, the visibility of FQH states in our experiments is sizeably reduced for larger optical excitation powers. This is due to two separate reasons. First, Rydberg excitons in the TMD monolayer get quenched upon increasing the power beyond 50\nW (see \figref{fig:technique_power_dep}{\bf a}). We find that this effect becomes more pronounced for higher-energy Rydberg states with larger principal quantum numbers, which is consistent with previous optical studies of Rydberg excitonic transitions in different material systems~\cite{Kazimierczuk2014}. 

Independently, the signatures of FQH states also become weaker for higher powers. This is revealed by \figref{fig:technique_power_dep}{\bf c} showing the \vtg-dependent $A_{4\mathrm{s}}$ measured at different excitation powers in the filling factor range between $2 \leq \nu\leq 3$. In order to compensate for the power-induced loss of $A_{4\mathrm{s}}$, the subsequent curves are normalized with respect to the height of the amplitude maximum at $\nu = 2$. We find that all of the visible FQH states become strongly suppressed even for powers as low as $\textapprox 30$\nW, which we attribute to finite graphene light absorption. In order to reduce the influence of both of the above-discussed issues, all of the experiments reported in the main text are performed using an excitation power of $\textapprox 9$\nW.


\begin{figure}[h!]
    \includegraphics[width=0.75\textwidth]{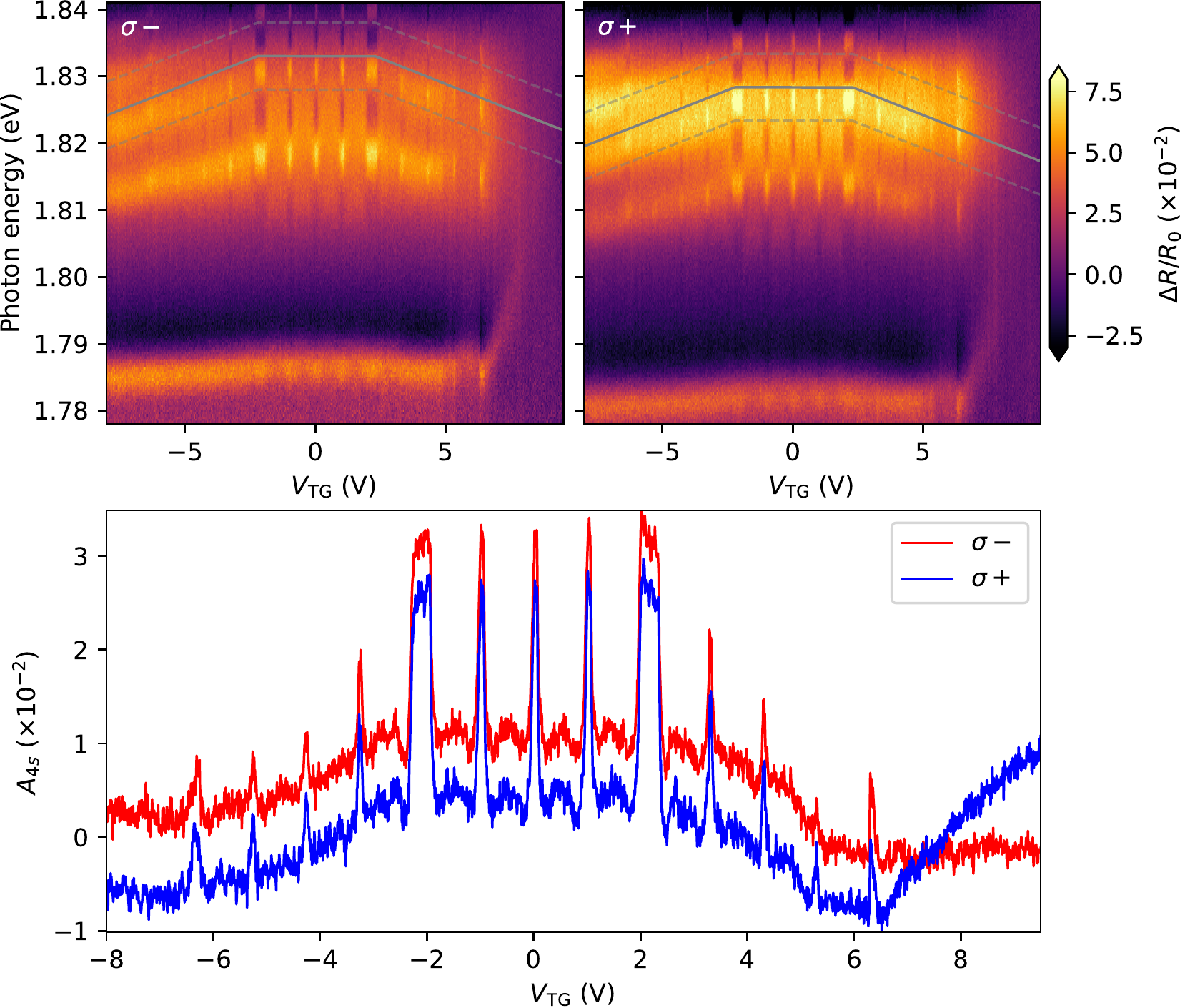}
    \caption{\label{fig:FQHE_both_pols_contrast}{\bf Optical response of Rydberg excitons in co-circularly-polarized excitation-detection basis.} Upper panels: \vtg-dependent \DRR\ spectra of Rydberg excitons acquired in the 5ML-spacer region at $B=16\tesla$, $T=4\kelvin$, and in $\sigma^-$ (left) and $\sigma^+$ (right) polarizations. Gray lines indicate the boundaries of the spectral integration regions used for extraction of $A_{4\mathrm{s}}$ shown in the lower panel. The integration regions are shifted with respect to each other between the two polarization by the valley-Zeeman splitting of $\approx4.6\meV$ corresponding to a g-factor of \textapprox 5~\cite{Goryca2019}. Lower panel: $A_{4\mathrm{s}}$ extracted from $\sigma^+$- and $\sigma^-$-polarized spectra.}
\end{figure}

\subsection{Circular-polarization-resolved measurement}

In all FQH measurements reported in the main text, we utilize a nearly cross-polarized excitation-collection basis to increase the reflectance contrast amplitude of the Rydberg optical transitions (see \secref{sec:pol_settings}). \figref{fig:FQHE_both_pols_contrast}{\bf a} shows the \vtg\ evolution of the \DRR\ spectra acquired at $B=16\tesla$ and $T=4\kelvin$ with co-polarized excitation and collection beams in either $\sigma^+$ or $\sigma^-$ polarizations. The \vtg-dependent amplitudes of the 4s exciton extracted based on these two datasets are shown in \figref{fig:FQHE_both_pols_contrast}{\bf b}. We do not observe any discernable differences between the optical responses of Rydberg excitons measured in the two circular polarizations. This observation justifies our approach of using nearly cross-linearly-polarized settings that combine the $\sigma^\pm$-polarized contributions to maximize the signal-to-noise ratio.

We note that the signatures of FQH states are not reliably observable in the circularly-resolved data. This is partially due to lower \DRR\ amplitude of the 4s state compared to that obtained in nearly cross-linearly-polarized settings in the main text. Furthermore, the measurements in circular basis were taken at higher temperature (4\kelvin) and with shorter integration time. 

\section{FQH measurements in different settings}
\label{sec:other_areas}

\begin{figure}[b!]
    \includegraphics[width=0.55\textwidth]{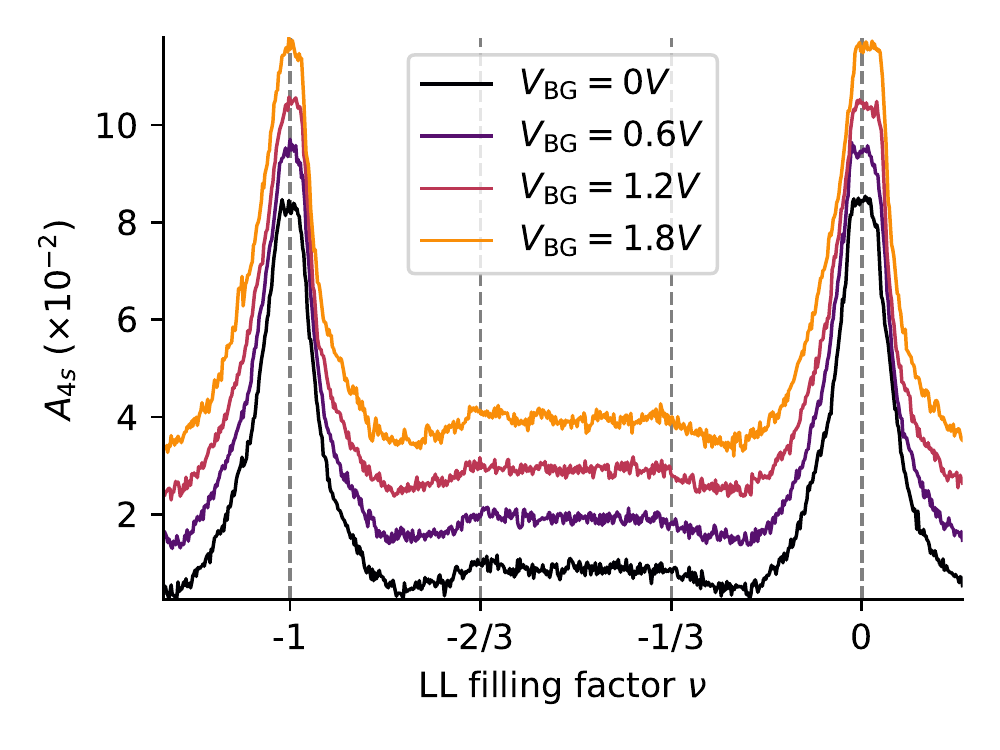}
    \caption{\label{fig:more_data_vbg_mlg}{\bf Independence of IQH and FQH signatures of the out-of-plane electric field.} Electron-density-dependent amplitude of the 4s exciton obtained for different $\vbg$ in the graphene-covered region at $B=16\tesla$ and $T=80\mkelvin$. In the measurement, the \vtg\ range is adjusted to compensate for the applied \vbg\ in order to ensure that the electron density in graphene is always varied in a range corresponding to $-1\leq \nu \leq 0$ (note that the curves are vertically offset for clarity). All the curves are almost identical. This confirms that, as expected, IQH and FQH states in graphene are independent of the value of the applied out-of-plane electronic field.}
\end{figure}

\subsection{Independence of FQH signatures in monolayer graphene of the electric field}

The data presented in the main text has been obtained without applying any voltage to the graphite back gate. In order to verify whether out-of-plane electric field has any effect on the analyzed FQH states, we have also measured top-gate-voltage-dependent amplitudes of the 4s exciton for different values of \vbg. The results of these experiments are presented in \figref{fig:more_data_vbg_mlg} as a function of graphene filling factor determined using the parallel-plate capacitor model introduced in \secref{sec:density_calib}. As expected for the monolayer graphene, all of the acquired curves lie on top of each other, which confirms that neither IQH nor FQH features in the graphene are influenced by the applied electric field. These findings also indicate that LL-quantization of the graphite gates has no effect on the investigated changes of the Rydberg exciton amplitudes.

\subsection{FQH signatures in other sample regions}

\begin{figure}[b!]
    \includegraphics[width=0.55\textwidth]{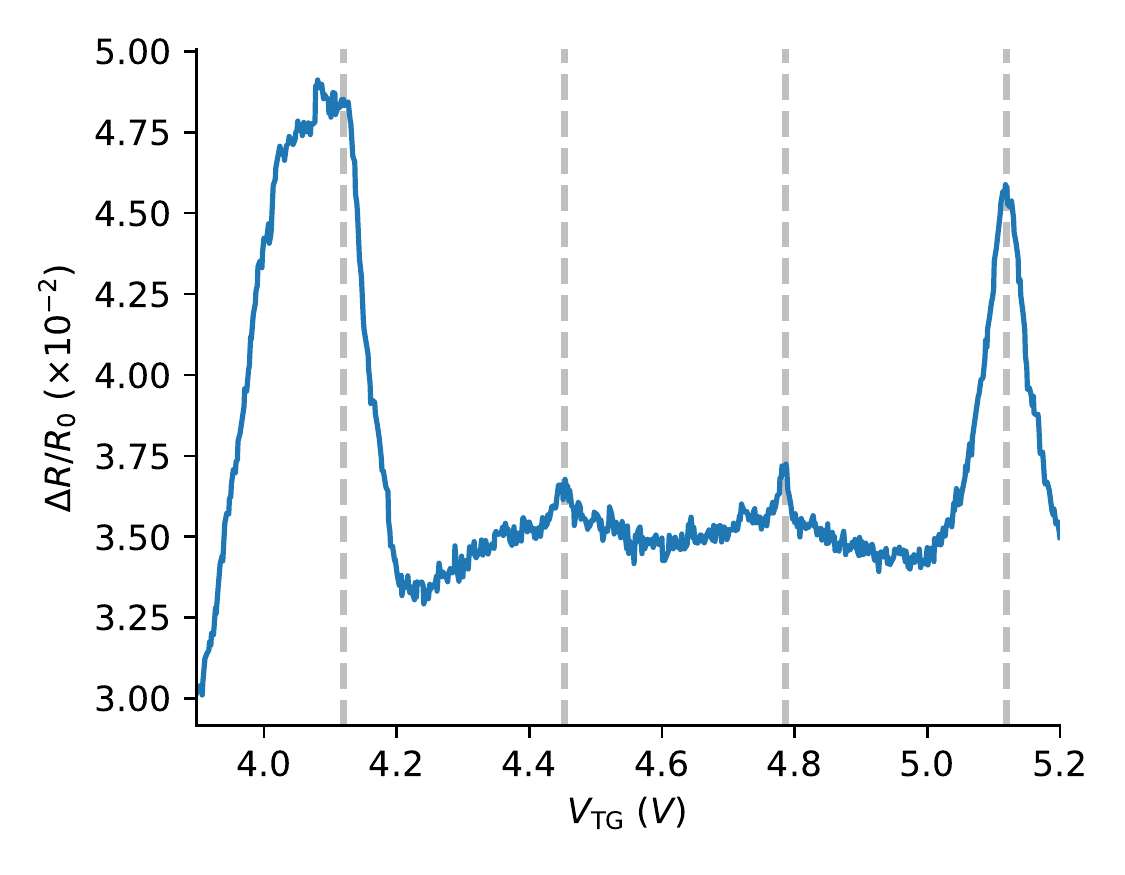}
    \caption{\label{fig:more_data_BLG_high_snr}{\bf Optical signatures of FQH states in bilayer graphene.} Amplitude of 4s Rydberg exciton determined based on $\vtg$-dependent reflectance contrast measurements performed in the bilayer-graphene region with a 5ML-thick hBN spacer at $B=16\tesla$, $T=80\mkelvin$, and for $4\leq \nu \leq 5$. Gray dashed lines indicate voltages corresponding to $n/3$ filling factors. The amplitude maxima at $4+1/3$ and $4+2/3$ fillings are clearly visible, evidencing the formation of FQH states.}
\end{figure}

All the experiments reported in the main text have been carried out in the graphene-covered region with a 5ML-thick hBN spacer, where Rydberg exciton transitions are comparatively strong. However, we have also utilized our optical sensing scheme to investigate the formation of FQH states in the BLG region featuring an hBN spacer of the same thickness. \figref{fig:more_data_BLG_high_snr} shows the optical measurement of the electronic compressibility in BLG carried out in this sample region at $B=16$~T and $T=80$~mK. The extracted $A_{4\mathrm{s}}$ exhibits the familiar enhancements at both integer and fractional filling factors, thus providing direct signatures of FQH states in BLG. Notably, the amplitude of these oscillations in $A_{4\mathrm{s}}$ is reduced by a factor of \textapprox 5 compared to the monolayer graphene region. This is a consequence of reduced amplitude of Rydberg excitons which are more efficiently screened due to higher density of states in BLG. 

As already noted in the main text, we have not been able to detect FQH states in yet another sample region with monolayer graphene separated from the TMD monolayer by a thinner, 2ML-thick hBN spacer. This is most likely due to markedly reduced amplitude of Rydberg excitons in this region even when the $E_F$ lies in the LL gap. We also speculate that the TMD monolayer, when placed in too close proximity, might have a detrimental effect on electronic correlations in graphene.